\documentclass[a4paper,11pt]{article}

\pdfoutput=1
\usepackage{comment}
\usepackage{fixltx2e}
\usepackage[table,xcdraw]{xcolor}
\usepackage{jcappub} 
\usepackage{float} 
\usepackage[scr=boondox]{mathalfa}
\usepackage{bm}
\usepackage{stmaryrd}
\usepackage{bbold}
\usepackage[normalem]{ulem}
\usepackage{amsmath}
\usepackage{bbm}

\def\be{\begin{equation}}
\def\ee{\end{equation}}
\def\bea{\begin{eqnarray}}
\def\eea{\end{eqnarray}}
\newcommand{\vs}{\nonumber\\}
\def\ba#1\ea{\begin{align}#1\end{align}}
\def\bg#1\eg{\begin{gather}#1\end{gather}}

\def\iMpch{\,h\,{\rm Mpc}^{-1}}

\newcommand{\s}{\sigma}

\newcommand{\refeq}[1]{Eq.~(\ref{eq:#1})}          
\newcommand{\refeqs}[2]{Eqs.~(\ref{eq:#1})--(\ref{eq:#2})}          
          
\newcommand{\reffig}[1]{Fig.~\ref{fig:#1}}
\newcommand{\reffigs}[2]{Figs.~\ref{fig:#1}--\ref{fig:#2}}
          
\newcommand{\reftab}[1]{Tab.~\ref{tab:#1}}          
\newcommand{\refsec}[1]{Sec.~\ref{sec:#1}}          
\newcommand{\refapp}[1]{App.~\ref{app:#1}}

\newcommand{\shellPT}[1]{{(#1)_{\rm shell}}}

\newcommand{\unit}{{\mathbb{1}}}

\def\Plin{P_{\rm L}}

\renewcommand{\v}[1]{\bm{#1}}

\renewcommand{\emph}[1]{\textit{#1}}

\newcommand{\vx}{\v{x}}

\newcommand{\vk}{\v{k}}
\newcommand{\vq}{\v{q}}

\newcommand{\vp}{\v{p}}

\newcommand{\<}{\langle}
\renewcommand{\>}{\rangle}

\DeclareMathOperator{\ii}{i}

\renewcommand{\d}{\delta}

\newcommand{\Z}{\mathcal{Z}}
\newcommand{\G}{\mathcal{G}}
\newcommand{\K}{\mathcal{K}}

\newcommand{\Shell}{\mathcal{S}}

\newcommand{\eps}{\epsilon}

\def\ddet{\delta_{\rm det}}

\def\L{\Lambda}

\def\leps{\lambda}

\def\dlin{\delta^{(1)}}
\def\dlinshell{\dlin_{\rm shell}}

\def\Zcorr{\zeta}

\def\L{\Lambda}

\def\P{\mathcal{P}}
\def\O{\mathcal{O}}
\def\o{p}
\def\Del{\mathcal{D}}

\newcommand{\perm}[1]{ \expandafter\ifstrempty\expandafter{#1} {\mbox{perm.}} {\mbox{$#1$ perm.}} }

\makeatletter
\newlength{\apb@width}
\newcommand{\autoparbox}[2][c]{\settowidth{\apb@width}{#2}\parbox[#1]{\apb@width}{#2}}
\newcommand{\includegraphicsbox}[2][]{\autoparbox{\includegraphics[#1]{#2}}}
\makeatother

\def\dirac{\delta_{\rm D}}
\def\diracpi{\hat{\delta}_{\rm D}}

\renewcommand{\comment}[1]{}

\usepackage[T1]{fontenc}
\usepackage{graphbox}
\usepackage{subfig}
\usepackage{booktabs}
\subheader{TUM-HEP-1509/24}
\title{The Renormalization Group for Large-Scale Structure: Origin of Galaxy Stochasticity}
\usepackage{amsmath}
\author[a]{Henrique Rubira,}
\author[b]{Fabian Schmidt}

\affiliation[a]{Physik Department T31, Technische Universit\"at M\"unchen,\\
James-Franck-Stra{\ss}e 1, D-85748 Garching, Germany}
\affiliation[b]{Max-Planck-Institut f\"{u}r Astrophysik,\\ 
Karl-Schwarzschild-Str. 1, 85748 Garching, Germany}

\emailAdd{henrique.rubira@tum.de}
\emailAdd{fabians@mpa-garching.mpg.de}

\abstract{
  The renormalization group equations for large-scale structure (RG-LSS) describe how the bias and stochastic (noise) parameters---both of matter and biased tracers such as galaxies---evolve as a function of the cutoff $\L$ of the effective field theory. In previous work, we derived the RG-LSS equations for the bias parameters using the Wilson-Polchinski framework. Here, we extend these results to include stochastic contributions, corresponding to terms in the effective action that are higher order in the current $J$.
  We derive the general local interaction terms that describe stochasticity at all orders in perturbations,
and a closed set of nonlinear RG equations for their coefficients.
    These imply that
a single nonlinear bias term generates all stochastic moments through RG evolution. Further, the
evolution is controlled by a different, lower scale than the nonlinear scale. This has implications for the optimal choice of the renormalization scale when comparing the theory with data to obtain cosmological constraints.} 

\keywords{Large-scale structure, galaxy clustering, bias, power spectrum, bispectrum, effective field theory, renormalization group}

\begin{document}

\maketitle
\flushbottom

\section{Introduction}\label{sec:intro}

Galaxy redshift surveys map out the large-scale structure (LSS) over a substantial part of the observable low-redshift universe, and constitute a rich source of information on gravity, dark matter, dark energy, and the initial conditions of structure formation. To unlock this information, however, one has to marginalize over the significant uncertainties in our understanding of galaxy formation. What is required in particular is a reliable prediction for the conditional probability of forming a galaxy that passes observational luminosity and color selections, at a given location. The effective field theory (EFT) approach \cite{Baumann:2010tm,carrasco/etal:2012,Carroll:2013oxa,Konstandin:2019bay,DAmico2019,Ivanov2019}, built upon cosmological perturbation theory, allows for a consistent expansion of the galaxy density field into operators $O$, ranked in terms of perturbative order, and corresponding bias coefficients $b_O$ as well as stochastic contributions. The latter can be effectively described by a field $\eps$ and coefficients $c_{\eps,O}$ (see \cite{Desjacques:2016bnm} for a review): 
\be
\d_g(\vx,\tau) \equiv \frac{n_g(\vx,\tau)}{\bar n_g(\tau)} - 1
= \sum_O \left[ b_O(\tau) + c_{\eps,O}(\tau) \eps(\vx,\tau) \right] O(\vx,\tau)
+ \eps(\vx,\tau) .
\label{eq:bias}
\ee
We emphasize that at the level of the partition function, there is no stochastic field $\eps$, as we will see below. Nevertheless, the field $\eps$ and its statistics are useful to give physical meaning to the stochastic terms appearing in the partition function.\footnote{Note that we only write a single field $\eps$ in \refeq{bias}, unlike the previous literature which has introduced multiple fields $\eps_O$. We will discuss this distinction in more detail in upcoming work.} 

When computing observable statistics such as $n$-point correlation functions based on \refeq{bias} (technically these involve expectation values of products of composite operators), loop integrals arise which have to be regularized. The standard approach has so far been to work at the level of $n$-point correlation functions, and to remove the UV-sensitive loop integrals via counterterms \cite{Assassi2014,Patrone:2023cqe}. In this paper, following up on \cite{Rubira:2023vzw} (henceforth ``Paper 1''), our aim is instead to pursue the renormalization-group (RG) approach of Wilson \cite{wilson:1971} and Polchinski \cite{polchinski:1984}, which proceeds by integrating out small-scale modes in the partition function, and which we refer to as ``RG-LSS'' in the following.
Ref. \cite{Carroll:2013oxa} was the first to point out how this approach can be adapted to the EFT of LSS. Ref.~\cite{Cabass:2019lqx} used it to derive the EFT prediction for the conditional probability density of the galaxy field given the matter density field; that is, the crucial ingredient needed for making predictions for galaxy clustering, as noted above. 

The Wilson-Polchinski approach to renormalization proceeds by integrating out modes in the free field above a momentum cutoff $\Lambda$.\footnote{The Wilson-Polchinski approach is often referred to as {\it nonperturbative} \cite{Dupuis:2020fhh,Berges:2000ew,Delamotte:2007pf}, in contrast to  Wilson's pioneering {\it perturbative} approach.} In the LSS case, this corresponds to integrating modes above the cutoff \emph{in the linear density field}.
Integrating out modes in a thin shell in momentum space leads to renormalization-group equations that govern the running of the coupling constants in the effective action, such as the bias coefficients $b_O$ (``RG flow'').
In Paper~1, we explicitly derived the RG equations governing the running of the bias coefficients, and showed that the basis of bias operators used there is closed under renormalization.
The bias coefficients describe the mean-field prediction for the galaxy density given a realization of large-scale perturbations. However, the small-scale modes that have been integrated out also lead to scatter around this mean-field relation, which we refer to as stochasticity. In this paper, we turn to these stochastic contributions to the galaxy density field, which are described by interaction terms in the effective action  
that involve second and higher powers of the current. We derive how nonlinear bias sources such stochastic contributions when integrating out modes, and describe the general hierarchy that governs the RG flow between different types of interactions in the effective action. 

We provide the most general RG running of {\it all} leading-in-derivative EFT parameters [see \refeq{masterSol}], elucidating in detail how bias coefficients source stochastic terms. This result could be used in practice as an extra validation (both at field level and using $n$-point correlation functions) for the running of the EFT parameters. In addition, our framework can straightforwardly renormalize any higher-order $n$-point functions (including their respective stochastic contributions). Moreover, our RG approach keeps the smoothing scale $\L$ explicit, going beyond the usual $n$-point function renormalization scheme considered in other works \cite{Assassi2014} and proven in Sec.~3 of Paper 1 to correspond the limit $\L\to 0$ of the RG equations. 

The outline of the paper is as follows. In the remainder of this section, we summarize the main results and define our notation. In \refsec{stoch_theory} we discuss the partition function that includes the most general set of stochastic operators as higher-order-in-$J$ terms. We derive in \refsec{running} the general RG flow for the terms in the bias expansion including stochasticity. \refsec{results} presents results for those RG equations and discusses their numerical solutions. We conclude in \refsec{conc}. \refapp{shell} is dedicated to the evaluation of the shell integrals, while \refapp{doubleshell} proves which shell diagrams need to be included in the RG flow.

\subsection{Summary of main results} \label{sec:introsummary}  

The main formal result of the paper is a general renormalization group equation (RGE)
for stochastic contributions to galaxy clustering. Stochasticity corresponds to contributions to the effective action that are higher-order in the current $J$, i.e. $\propto J^m$ with $m>1$. The general effective action for the galaxy density field is given in \refeq{Sreal}
\be \label{eq:Sreal_int}
S_{\rm eff}[\dlin_\L, J_\L] = \sum_{m=1} \sum_O \frac{C^{(m)}_{O}(\L)}{m!} \int_{\vx} \left[J_\L(\vx)\right]^m O[\dlin_\L](\vx)\,.
\ee
Note that the set of operators $O$ appearing here is always the same, and terms with $m=1$ are nothing but the bias expansion with $C_O^{(1)} \equiv b_O$. We explain how $S_{\rm eff}$ generates the $n$-point functions of galaxy clustering after \refeq{Zderiv}. 
Expanding the operator functionals $O[\dlin_\L]$ in \refeq{Sreal_int} [as discussed in more detail below, cf. \refeq{Ok}], we recover the same structure as the effective action introduced by \cite{Carroll:2013oxa}
\be
S_{\rm eff}[\dlin_\L, J_\L] = \sum_{m=1}^\infty\sum_{n=0}^\infty \frac1{n!m!}\K^{(m) j_1\ldots j_m}_{i_1\ldots i_n}(\L) (\dlin_\L)^{i_1}\cdots (\dlin_\L)^{i_n}\; J^\L_{j_1}\ldots J^\L_{j_m},
\label{eq:S_expanded}
\ee
with each $\K^{(m) j_1\ldots j_m}_{i_1\ldots i_n}(\L)$
representing a sum over operator kernels with associated coefficients $C_O^{(m)}$.\footnote{Specifically, we have 
  \be
  \K^{(m) j_1\ldots j_m}_{i_1\ldots i_n}(\L) = (2\pi)^3 \dirac(\vk_{j_1 \ldots j_m} + \vk_{i_1\ldots i_n}) \sum_O C^{(m)}_O (\L) K^{(n)}_{O}(\vk_{i_1}, \ldots, \vk_{i_n}),
  \ee
  where the sum runs over all bias operators contributing at order $n$ and $K^{(n)}_O$ is defined in \refeq{Ok}.
  The fact that Ref.~\cite{Carroll:2013oxa} considered a special case of biased tracer, matter, is not important for the structure of the equations.}
Ref.~\cite{Carroll:2013oxa} derive RG equations for the coefficient
tensors $\K^{(m)}(\L)$ [Eq.~(4.22) there; notice that in their notation, $n$ and $m$ are swapped and their $K$ corresponds to our $\K$]. For clarity, let us drop all indices and combinatorial factors in the following.
The structure of the RGE derived by Ref.~\cite{Carroll:2013oxa} is
\ba
    \frac{d}{d\L} \K^{(m)}(\L) &=  \frac{d\Plin^{\L}}{d\L} \cdot \K^{(m)}
+ \sum_{\substack{m_1=1, m_2=1 \\ m_1+ m_2 = m}}^m \K^{(m_1)} \cdot \frac{d\Plin^{\L}}{d\L} \cdot \K^{(m_2)}  \vs &= \raisebox{-0.0cm}{\includegraphicsbox[scale=0.7]{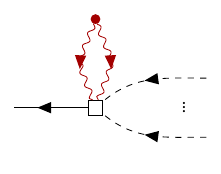}} + \raisebox{-0.0cm}{\includegraphicsbox[scale=0.7]{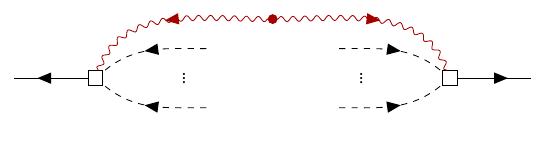}} \,, \label{eq:RGECarroll}
\ea
where $d\Plin^{\L}/d\L$, the derivative of the cut linear power spectrum with respect to the cutoff, stands for the inverse propagator, or covariance of linear modes in the infinitesimal momentum shell $[\L,\L+\lambda]$, and $\cdot$ denotes index contractions, i.e. momentum integrals. 
In the lower line of \refeq{RGECarroll}, we have written the corresponding Feynman diagrams illustrating the modes in the momentum shell that are being integrated out (red wiggly lines; for our complete Feynman conventions, see \refsec{notation}).

The two contributions in \refeq{RGECarroll} are precisely the two terms identified in the original paper by Polchinski \cite{polchinski:1984}, which each involve precisely one propagator in the momentum shell. The downside that comes with this simple structure is that one obtains nonlocal interactions. While these can be directly expanded in derivatives when considering the first ``loop'' term in \refeq{RGECarroll}, the second ``tree-level'' cannot be directly expanded. 
  In the simplest nontrivial case for our effective action, this interaction looks like [\refeq{S11_appB}]
  \be
  \int_{\vx} J_\L(\vx) \dlin_\L(\vx) 
\int_{\vx'} J_\L(\vx') \dlin_\L(\vx') \xi_{\rm shell}(\vx-\vx'),
\ee
where $\xi_{\rm shell}(r)$ is the correlation function of modes in the momentum shell (i.e. the Fourier transform of the shell propagator).
This type of nonlocal term can be shown to vanish in the limit that the external momenta contributing to the current $J_\Lambda$ are much less than the scale $\L$ (\refapp{J2structure}); however, in the Polchinski approach we cannot directly set the external momenta to be small, because they include the momentum shell that is to be integrated out in the  next step.

The question then is, whether and if so, how, the local interactions in \refeq{Sreal_int} arise. We show in this paper that it is possible to work with soft external momenta in the RG flow from the beginning, resulting in local interactions from the outset. The price to pay for this significant simplification is that one now needs to consider contributions with higher powers of shell propagators in the RG flow. Nevertheless, we succeed in arriving at a closed set of (nonlinear) RGE, the \emph{general local RG-LSS equations} in the form of
\ba
\frac{d}{d\L} \K^{(m)}(\L) &=  \frac{d\Plin^{\L}}{d\L} \cdot \K^{(m)}
+ \sum_{\substack{m_1=1, m_2=1 \\ m_1+ m_2 = m}}^m \frac{d\Plin^{\L}}{d\L} \cdot \K^{(m_1)} \cdot \frac{d\Plin^{\L}}{d\L} \cdot \K^{(m_2)} 
\vs
&\hspace*{1cm} + \sum_{\substack{m_1=1, m_2=1, m_3=1 \\ m_1+ m_2+m_3 = m}}^m \frac{d\Plin^{\L}}{d\L} \cdot \K^{(m_1)} \cdot \frac{d\Plin^{\L}}{d\L} \cdot \K^{(m_2)} \cdot \frac{d\Plin^{\L}}{d\L} \cdot \K^{(m_3)}
+ \ldots \vs
& = \raisebox{-0.0cm}{\includegraphicsbox[scale=0.7]{figs/diag_S2_nolabels.pdf}} + \raisebox{-0.0cm}{\includegraphicsbox[scale=0.7]{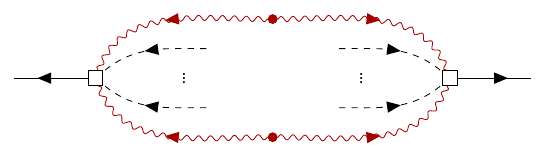}} 
\vs  
&\quad + \raisebox{-0.0cm}{\includegraphicsbox[scale=0.6]{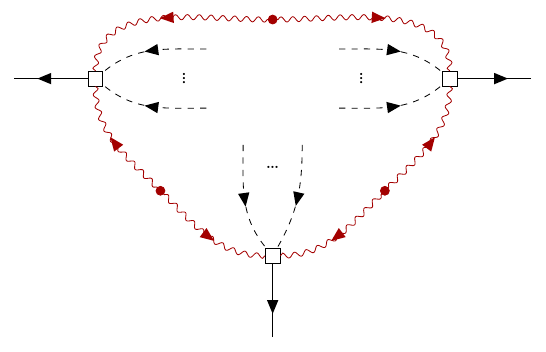}} + \ldots + \raisebox{-0.0cm}{\includegraphicsbox[scale=0.6]{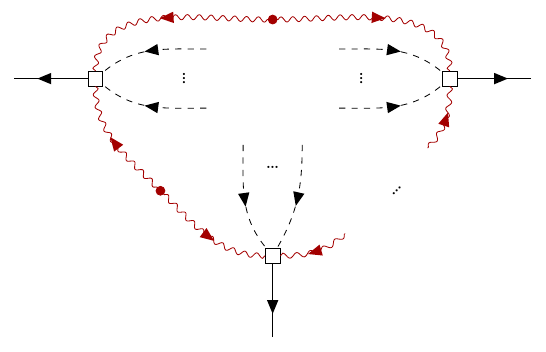}} \,, \label{eq:RGEgeneral}
\ea
where the series continues to all orders in $d\Plin^\L/d\L$. While one might think that terms with higher powers of $d\Plin^\L/d\L$ are suppressed since one is considering a narrow shell in momentum space, this is not the case as we will show; conversely, all other possible types of diagrams are suppressed (the reason: the terms in \refeq{RGEgeneral} are all \emph{one-loop diagrams}).
Note that for $m=1$ (bias terms), only the first term in both \refeq{RGECarroll} and \refeq{RGEgeneral} contributes, and we recover the RGE for bias parameters studied in detail in Paper 1.

Clearly, the structure of \refeq{RGEgeneral} is quite different from, and more complex than that of \refeq{RGECarroll}.
However, unlike \refeq{RGECarroll}, \emph{we can work with local interactions from the beginning.}
We will explain in detail how \refeq{RGEgeneral} arises, and what the physical ramifications are. The \textbf{main results} are the following:
\begin{itemize}
\item \refeq{RGEgeneral} implies that, under RG flow, a single nonlinear bias term such as $b_{\delta^2}$ immediately generates terms of any order $J^m$ in the current; that is, it generates \emph{all moments of stochasticity}.
\item The finding that terms of order $J^m$ in the effective action do not source terms $J^{m'}$ with $m' < m$ (in particular, stochasticity does not source bias terms) remains valid.
\item Despite the seemingly complicated structure of \refeq{RGEgeneral}, we are able to derive an explicit closed form valid at any order in $m$ [\refeq{masterSol}]. This is possible thanks to the specific, simple structure of the diagrams involved: going to higher order in $m$ amounts to inserting another factor of $d\Plin^\L/d\L \cdot \K^{(m_j)} \cdot$ given by lower-order kernels $m_j < m$.
\item Finally, we express \refeq{RGEgeneral} as an explicit set of nonlinear ordinary differential equations for the stochastic amplitudes and bias parameters $C_O^{(m)}(\L)$.
  This nonlinearity leads to interesting behavior under RG flow, and ramifications for the choice of the renormalization scale in EFT analyses,  which we investigate in \refsec{results}.
\end{itemize}
It appears much more difficult to arrive at such explicit results when dealing with the Polchinski RG \refeq{RGECarroll} with nonlocal interactions instead.

In short, we present the most general RG running of all EFT parameters [see \refeq{masterSol}], explaining how stochastic terms are sourced via the running of the EFT smoothing scale and showing that the effective action \refeq{Sreal_int} is closed under renormalization. 

\subsection{Notation}
\label{sec:notation}

We follow a similar notation to Paper~1, which we review here.

For the Fourier conventions, we use $\vk,\vp$ and $\vq$ for momenta variables where bold letters represent three-vectors and $\vp_{1 \dots n} = \vp_{1} + \dots + \vp_{n}$ as a short notation for the sum of vectors. 
We also use prime in the subscript or in the variable interchangeably, e.g. $\vp_{1}' = \vp_{1'}$.
Fourier-space integrals are written as
\be
\int_{\vp_1,\dots,\vp_n} = \int \frac{d^3p_1}{(2\pi)^3} \dots \int \frac{d^3p_n}{(2\pi)^3} \,.
\ee
The corresponding real-space normalization is
\be
\int_{\vx_1, \dots, \vx_n} = \int d^3 x_1\, \dots \int d^3 x_n   \,.
\ee
We denote fields smoothed with a sharp-k filter $W$ on a length scale $1/\L$ as $f_\L$, where, in Fourier space,
    \be
        f_{\L}(\vk) = \,W_\L(\vk) f(\vk)\,.
    \ee

We define an operator $O[\d]$ of order $n$ as a function of $n$ insertions of the matter overdensity $\d$
    \be
    O[\d](\vk) = \int_{\vp_1,\ldots,\vp_n} \diracpi(\vk-\vp_{1\ldots n}) S_O(\vp_1,\ldots \vp_n) \d(\vp_1) \cdots \d(\vp_n).
    \label{eq:Odef}
    \ee
Here we adopt $\diracpi = (2\pi)^3 \dirac$ for the Dirac delta.
We also use as the bias basis of operators
    \ba
    \mbox{Zeroth order:}&\quad \unit \,;\vs
    \mbox{First order:}&\quad \d \,;\vs
    \mbox{Second order:}&\quad \d^2,\  \G_2\,; \vs
    \mbox{Third order:}&\quad \d^3,\  \d\,\G_2,\  \Gamma_3,\  \G_3 \,.
    \ea
    Notice that we included the zeroth-order unit operator $\unit$, which starts to contribute at order $J^2$ since the $J^1$ term leads to a tadpole contribution that is subtracted after setting $\langle \d_g \rangle = 0$. In Fourier space, the unit operator is a Dirac delta.
    The Galileon operators are defined as 
    \bea
    \G_2( \Phi_g) &\equiv& (\nabla_i\nabla_j\Phi_g)^2 - (\nabla^2  \Phi_g)^2 \label{eq:galileon2}\,, \\
    \G_3( \Phi_g) &\equiv& -\frac{1}{2}\left[2\nabla_i\nabla_j  \Phi_g \nabla^j\nabla_k  \Phi_g\nabla^k\nabla^i  \Phi_g + (\nabla^2  \Phi_g)^3-3(\nabla_{i}\nabla_{j} \Phi_g)^2\nabla^2  \Phi_g\right]\,, \label{eq:galileon3}
    \eea
    with $\Phi_g \equiv \nabla^{-2}\d$ being the scaled gravitational potential. At third order we have
    \bea
    \Gamma_3( \Phi_g,\Phi_v) \equiv \G_2(\Phi_g) - \G_2(\Phi_v) \,,
    \eea
    that also contains $\Phi_v \equiv \nabla^{-2} \bm{\nabla}\cdot\v{v}$, the  velocity potential. Moreover, we define
    \be
    S_{\G_2} =  \s^2_{\vk_1,\vk_2}  = \left( \vk_1 \cdot \vk_2/k_1k_2 \right)^2 - 1 \,.
    \ee 
    In order to avoid tadpole contributions, we normalize (starting from first-order operators)
    \be \label{eq:Onorm}
    O(\vk) \to O(\vk) - \langle O(\vk=0)\rangle \,.
    \ee
    
    We further expand the matter density field as
    \be
    \d(\vk) = \sum_{\ell=1}^\infty\int_{\vp_1,\ldots,\vp_\ell} \diracpi(\vk-\vp_{1\ldots \ell}) F_\ell(\vp_1,\ldots \vp_\ell) \dlin(\vp_1) \cdots \dlin(\vp_\ell),
    \label{eq:dexp}
    \ee
    where $F_\ell$ are the usual PT kernels (with $F_1 \equiv 1$) \cite{lssreview} such that
    \be
    O(\vk) = \sum_{\ell={\rm order}(O)}^\infty\int_{\vp_1,\ldots,\vp_\ell} \diracpi(\vk-\vp_{1\ldots \ell}) K_O^{(\ell)}(\vp_1,\ldots \vp_\ell) \dlin(\vp_1) \cdots \dlin(\vp_\ell) \,.
    \label{eq:Ok}
    \ee
    One can derive the $K^\ell_O$ by inserting \refeq{dexp} into \refeq{Odef}. We also use $K_O \sum_{\ell={\rm order}(O)}^\infty K^{(\ell)}_O$. Therefore, we can more generically write $O[\d[\d_\L^{(1)}]]$ or simply the shorthand notation $O[\d_\L^{(1)}]$.
    The three types of vertices are represented by
    \be
    \raisebox{-0.0cm}{\includegraphicsbox[scale=0.9]{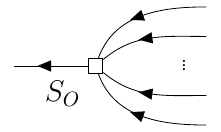}}\qquad
    \raisebox{-0.0cm}{\includegraphicsbox[scale=0.9]{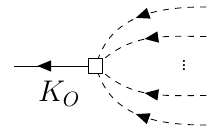}}\qquad
    \raisebox{-0.0cm}{\includegraphicsbox[scale=0.9]{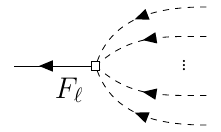}}
    \quad,
    \ee
    where the large boxes indicate the operator convolution \refeq{Odef} or \refeq{Ok} and the small boxes represent the expansion of $\d$, \refeq{dexp}. The linear density-field legs are represented as dashed lines.

We write down the partition function
    \be \label{eq:Z_intro}  
        \Z [J_\L] = \int \Del\dlin_\L \, \P[\dlin_\L] \exp\left(S_{\rm eff}[\dlin_\L,J_\L]  \right)\,,
    \ee
where $S_{\rm eff}[\dlin_\L,J_\L]$ is the effective action (see \refsec{stoch_theory}) and
    \be
        \P[\dlin_\L] = \left(\prod_{\vk}^\L 2\pi \Plin(k)\right)^{-1/2} \exp\left[-\frac12 \int_{\vk}^\L \frac{|\d^{(1)}|^2}{\Plin(k)}\right]\,, \label{eq:PlikeDef}
    \ee
the probability distribution function (PDF) of the Gaussian field $\dlin_\L$. 

We consider two types of propagators: the linear propagator cut at $\L$ 
    \be
    \Plin^\L(k) = \< \d^{(1)}_\L(\vk)\d^{(1)}_\L(\vk')\>^{'},
    \ee
    and the shell propagator that only has support within an infinitesimal shell $[\L,\L+\leps]$ in momentum space
    \be \label{eq:Pshell}
    P_{\rm shell}(k) = \< \dlinshell(\vk)\dlinshell(\vk')\>^{'} \,.
    \ee
    Those propagators are represented as
    \ba
    \qquad \raisebox{-0.0cm}{\includegraphicsbox[scale=0.8]{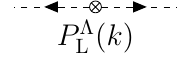}} \qquad
    \raisebox{-0.0cm}{\includegraphicsbox[scale=0.8]{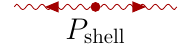}} . \nonumber
    \ea
    We also use the shorthand prime notation 
      \be
      \<O(\vk_1) \dots O(\vk_n)\> = \, \diracpi(\vk_{1\dots n})\,\<O(\vk_1) \dots O(\vk_n)\>'\,.
      \ee
    Finally, the variance of the linear density field is defined as
    \be
    \sigma^2_\L  = \int_{\vp}^\L \Plin(p) .
    \label{eq:sigmaLambda}
    \ee  

    For the effective stochastic field, we can define the leading stochastic contribution to $m$-point functions as 
    \ba
    \langle \eps(\vk_1) \eps(\vk_2) \rangle &= \diracpi(\vk_{12}) P_{\eps,\unit}\,, \\
    \langle \eps(\vk_1)  \eps(\vk_2) \eps(\vk_3) \rangle &= \diracpi(\vk_{123}) B_{\eps,\unit}\,, \\
    \langle \eps(\vk_1)  \dots \eps(\vk_m) \rangle &= \diracpi(\vk_{1\dots m}) C^{(m)}_{\eps,\unit} \,.
    \label{eq:Cmunit}
    \ea
    The higher-derivative stochasticity can be defined in real space as, for instance for $P_{\eps,\nabla^2\unit}$,
    \be
    \langle \eps(\vx) \eps(\v{y}) \rangle = [P_{\eps,\unit} + P_{\eps,\nabla^2\unit} \nabla_{\vx}^2 + \ldots] \dirac(\vx-\v{y}).
    \ee
    We will discuss this further in the next section.
    The stochastic cross-moment involving an operator $O$ can be defined as, for instance
    \ba
    \langle \eps(\vk_1)  \eps(\vk_2) O(\vk_3) \rangle &= \diracpi(\vk_{123}) P_{\eps,O} \, O(\vk_3)\,,
    \ea
    or more generally 
    \ba
    \langle \eps(\vk_1) \dots  \eps(\vk_m) O(\vk_{m+1}) \rangle &= \diracpi(\vk_{1\dots m}) C_{\eps,O}^{(m)} \, O(\vk_{m+1})\,.
    \ea
    As an important point here, notice that any correlator involving a single instance of $\eps$ vanishes, i.e. $\eps$ does not correlate linearly with any of the other operators:
    \ba
    \langle \eps(\vk_1) O(\vk_2) O'(\vk_3) \ldots \rangle &= 0 \,.
    \ea
We stress again that these relations mainly serve to give meaning to the coefficients $P_{\eps,\unit}, P_{\eps, O}$ and $C_{\eps,O}^{(m)}$ in general; the partition function is completely written in terms of these coefficients and without reference to $\eps(\vk)$. An important result we obtain is that the action \refeq{Sreal}, including stochasticity, is closed under renormalization. Therefore the treatment of stochasticity that we provide in this work covers its most general properties, regardless of the LSS tracer considered, and also directly determines the stochastic contributions to all $n$-point functions.

We use a Planck 2018 Euclidean $\Lambda$CDM cosmology  \cite{Planck:2018vyg} for all numerical results.

\section{The general EFT partition function for bias and stochastic parameters} \label{sec:stoch_theory}

The focus of this section is to generalize the partition function of Paper 1,
\ba
\Z[J_{\L}] = \int \Del\dlin_\L \P[\dlin_\L] \exp& \left( S_{\rm eff}[\dlin_\L, J_\L] \right)
\vs 
= \int \Del\dlin_\L \P[\dlin_\L] \exp&\left(\int_{\vk} J_{\L}(\vk) \left[\sum_O b_O(\L) O[\dlin_\L](-\vk)\right]  \right. \vs
&\quad \left. + \frac12 P_{\eps,\unit}(\L) \int_{\vk} J_{\L}(\vk) J_{\L}(-\vk)  + \O[J_\L^2 \dlin_\L,\  J_\L^3]  \right)\,, \label{eq:ZofJ} 
\ea
beyond the leading stochastic contributions considered there. Here,
the sum runs over all bias operators $O$ that are relevant at a given order in perturbations and derivatives.\footnote{For other works on the large-scale structure partition function, see also \cite{matarrese/pietroni,floerchinger/etal,Blas:2015qsi,Blas:2016sfa,Cabass:2019lqx}.} 
Notice that \refeq{ZofJ} includes a nonlinear coupling between the field $\dlin_\L$ and the current $J$, via the operators $O[\dlin_\L]$. Terms that couple nonlinearly to the current are usually referred to as ``composite operators'' (see \cite{Shore:1990wp,Polonyi:2000fv} and the books \cite{Zinn-Justin:2002ecy, Collins:1984xc}). As discussed in \cite{Shore:1990wp}, the renormalization of those composite operators will lead to terms that are higher-order in $J$.
As we will see in \refsec{running}, those higher-order stochastic contributions are necessary to obtain a partition function that remains consistent under the RG flow.
Thus, the general effective action, writing explicitly the $\O[J_\L^2 \dlin_\L,\  J_\L^3]$ terms of \refeq{ZofJ}, is given by a sum over local-in-space products of currents and bias operators
\be \label{eq:Sreal}
\boxed{
S_{\rm eff}[\dlin_\L, J_\L] = \sum_{m=1} \sum_O \frac{C^{(m)}_{O}(\L)}{m!} \int_{\vx} \left[J_\L(\vx)\right]^m O[\dlin_\L](\vx)\,,}
\ee
where we include the constant operator $O(\vx) = \unit$ in the set of bias operators, which is more convenient when going to higher order in $J$. We can then identify $b_O(\L) \equiv C^{(1)}_{O}(\L)$ as the bias parameters\footnote{Notice that the tadpole coefficient $b_\unit = C^{(1)}_{\unit} = 0$ after setting $\langle \d_g \rangle = 0$.} at the scale $\L$, and $P_{\eps,O}(\L) \equiv C^{(2)}_{O}(\L)$ as the second-order ``Gaussian'' stochastic amplitudes, with $P_{\eps, \unit}$ corresponding to the well-known stochastic or ``shot noise'' contribution to the galaxy power spectrum. Similarly, $P_{\eps,\d}$ quantifies the leading coupling of stochasticity to the density (``density-dependent shot noise''). However, \refeq{Sreal} also includes non-Gaussian stochastic contributions, starting at $m=3$ with $B_{\eps,\unit} \equiv C^{(3)}_{\unit}$ as the purely stochastic contribution to the galaxy bispectrum.

Notice that the interactions among the $J$ and $O$ in \refeq{Sreal} are purely local. As noted in \refsec{introsummary}, the fact that we can write a local expression for the partition function that is closed under RG flow is a nontrivial result. Exact locality in fact is only obtained when working at leading order in derivatives for the stochastic contributions. Higher-derivative contributions, such as $J^{m-1} \nabla^2 J$ and $J^{m-2}\partial_iJ \partial^i J$ are also present in general, and are expected to be controlled by the same scale as that controlling the higher-derivative bias operators. We will neglect them in the following, however.

It is useful to study the dimension of each coefficient $C^{(m)}_{O}$. Let the dimension of a given $O$ be $d_O$; for an operator that is leading-order in derivatives\footnote{We count the number of derivative with respect to the density field $\d$. E.g., $\G_2$ has $d_{\G_2}=0$, while $\nabla^2 \d$ has $d_{\nabla^2 \d} = 2$.} we have $d_O=0$, while higher-derivative operators will have $d_O = n_{\rm deriv}(O)$. Then, by dimensional analysis
\ba
[O(\vx)] &= d_O \quad\Rightarrow\quad  [O(\vk)] = d_O-3\,, \\
[J(\vx)] &= 3 \quad\Rightarrow\quad  [J(\vk)] = 0\,, \\
[C^{(m)}_{O}] &= - \left[\int_{\vx} J^m O \right] = 3-3m-d_O \label{eq:dimC}\,.
\ea

The Fourier-space $m$-point correlation function of a biased tracer is given by 
\ba
\langle \d_g(\vk_1) \dots \d_g(\vk_m) \rangle &= \int \Del\dlin_\L \P[\dlin_\L] \, \d_g(\vk_1) \dots \d_g(\vk_m) \, e^{S_{\rm eff}} \,,
\ea
which can be obtained by taking derivatives of the partition function with respect to $J$, evaluated at $J=0$:
\be
\langle \d_g(\vk_1) \dots \d_g(\vk_m) \rangle =   \frac{\delta^m \Z }{\delta J(\vk_1) \dots \delta J(\vk_m)}\Big|_{J=0} \,.
\label{eq:Zderiv}
\ee
We find for instance
\ba
\langle \d_g(\vk) \rangle &= C^{(1)}_\unit = 0  \,,\\
\langle \d_g(\vk_1) \d_g(\vk_2) \rangle &= \sum_O \sum_{O'} C^{(1)}_O C^{(1)}_{O'} \langle O(\vk_1) O'(\vk_2) \rangle + C^{(2)}_\unit  \,, \label{eq:dd_general}\\
\langle \d_g(\vk_1) \d_g(\vk_2) \d_g(\vk_3) \rangle &= \sum_O \sum_{O'} \sum_{O''} C^{(1)}_O C^{(1)}_{O'}C^{(1)}_{O''} \langle O(\vk_1) O'(\vk_2)O''(\vk_3) \rangle \label{eq:bispec} \\ 
&\quad+ \left( \sum_O \sum_{O'} C^{(1)}_O C^{(2)}_{O'} \langle O(\vk_1) O'(\vk_{23}) \rangle + 2\,\textrm{perm.} \right)+  C^{(3)}_\unit \,,\nonumber
\ea
where we have used \refeq{Onorm} for all operators apart from $\unit$.
This structure continues similarly toward higher order.
Notice that these expressions are valid at any loop order in the perturbative expansion in $\dlin$.
Thus, from the structure of the $n$-point functions, we can read off that the $C^{(m)}_\unit$ corresponds to the $m$-th moment of the (purely) stochastic contribution to the galaxy density field. This term first appears in the $m$-point function as a $k$-independent contribution, whereas the other $C^{(m)}_O$ coefficients appear in $n$-point functions with $n>m$.

\section{RG flow via the partition function} \label{sec:running}
In this section, we derive the RG flow equations based on the Wilson-Polchinski formalism, which explicitly integrates out high-momentum modes in the partition function. We
generalize the bias RG equations developed in Paper 1 to include stochastic parameters for all higher $n$-point functions. We discuss the general structure of the running of the bias and stochastic parameters for all $n$-point functions. The reader interested in the main conclusions of this section can skip directly to \refsec{summary}.

\subsection{Integrating out a momentum shell via Wilson-Polchinski} \label{sec:WP}

In this section we follow the same procedure described by \cite{Carroll:2013oxa,Rubira:2023vzw}, splitting the integration functional between two cutoffs $\L$ and $\L'$
\be \label{eq:shelldef}
\dlin_{\L'}(\vk) = \dlin_\L(\vk) + \dlinshell(\vk),
\ee
in which $\dlinshell(\vk)$ is a field that has support only in an infinitesimal shell of width $\lambda = \L' - \L$. The partition function, written in real space, becomes
\ba
&\Z[J_{\L}] = \int \Del\dlin_\L \P[\dlin_\L] \int \Del\dlinshell \P[\dlinshell]  \\
&\hspace{3cm} \times\exp\left(\sum_m \frac{1}{m!}\int_{\vx}  \left[ J_{\L}(\vx) \right]^m  \sum_O C_O^{(m)}({\L'}) O[\dlin_\L+\dlinshell](\vx)  \right) \,, \nonumber
\ea
where $J_{\L}(\vx)$ is the real-space current, i.e., the Fourier transform of $J_{\L}$. Notice that we chose the current $J$ to have support only at $\L$ and not $\L'$ and therefore the partition function on the LHS also has support only up to $\L$. This is essential in the derivation, as it guarantees that the current is orthogonal to $\dlinshell$,
\be \label{eq:orthocondition}
\int_{\vk}J_\L(\vk)\dlinshell(-\vk) = 0\,.
\ee
This of course still allows for the evaluation of the partition function (and therefore the $n$-point functions derived from it) at momenta lower than $\L$ (see Paper 1 for a broader discussion). 
We can then expand the operator $O^{(n)}[\dlin_\L+\dlinshell]$, that is, $O$ evaluated at $n$-th order in perturbation theory, into contributions with different powers of $\dlinshell$:\footnote{For instance, the expansion of $\d[\dlin_\L+\dlinshell]$ can be written as
\ba
\d[\dlin_\L+\dlinshell](\vk) &=  \d[\dlin_\L](\vk) + \dlinshell(\vk) \\
 &+ \int_{\vp_1,\vp_2} \diracpi(\vk-\vp_{12}) F_2(\vp_1,\vp_2) \left[ 2\dlinshell(\vp_1)\dlin_\L(\vp_2) + \dlinshell(\vp_1)\dlinshell(\vp_2) \right] \vs
&+ \int_{\vp_1,\vp_2,\vp_3} \diracpi(\vk-\vp_{123})\, F_3(\vp_1,\vp_2,\vp_3) \left[ 3\,\dlinshell(\vp_1)\dlin_\L(\vp_2)\dlin_\L(\vp_3)\right. \vs
&\hspace{1cm} \left.+ 3\,\dlinshell(\vp_1)\dlinshell(\vp_2)\dlin_\L(\vp_3) + \dlinshell(\vp_1)\dlinshell(\vp_2)\dlinshell(\vp_3)\right] + \O\left[\left(\d^{(1)}\right)^4 \right]\,. \nonumber
\ea
For other examples of this expansion, see Appendix~A.1 of Paper 1.}
\bea \label{eq:Oshell}
O^{(n)}[\dlin_\L+\dlinshell] &=& O^{(n)}[\dlin_\L] + O^{(n),\shellPT{1}}[\dlin_\L, \dlinshell] + O^{(n),\shellPT{2}}[\dlin_\L, \dlinshell] \\
&& \qquad\qquad\qquad \qquad  \qquad + \ldots + O^{(n),\shellPT{n-1}}[\dlin_\L, \dlinshell] + O^{(n)}[\dlinshell]\,. \nonumber
\eea
After integrating out the momentum shell, we find
\ba
&\Z[J_{\L}] = 
\int \Del\dlin_\L \P[\dlin_\L] \exp\left(\sum_m \bigg\{\frac{1}{m!} \int_{\vx} \left[ \left( J_{\L}(\vx) \right)^m \sum_O C_O^{(m)}({\L'}) O[\dlin_\L](\vx)\right] +  \Zcorr^{(m)}[J_\L, \dlin_\L]\bigg\} \right)  \,,
\ea
where each of the corrections from the shell modes, $\Zcorr^{(m)}[J_\L, \dlin_\L]$ is of order $J^m$.
The $m=1$ correction is given by
\ba
& \Zcorr^{(1)}[J_\L, \dlin_\L] = \int_{\vx} J_{\L}(\vx) \sum_O b_O({\L'})  \Shell_{O} [\dlin_\L](\vx) \,, \label{eq:corr1}
\ea
as already pointed out in Paper 1. The $m=2$ correction, $O(J^2)$, is given by
\ba
& \Zcorr^{(2)}[J_\L, \dlin_\L] = \frac12 \int_{\vx}  \left[ J_{\L}(\vx) \right]^2  \sum_{O} P_{\eps,O}({\L'})   \Shell_{O} [\dlin_\L](\vx)
 \label{eq:corr2} \\
&\hspace{2cm} + \frac12 \int_{\vx_1,\vx_2} J_{\L}(\vx_1) J_{\L}(\vx_2)   \sum_{O, O'} b_O({\L'}) b_{O'}({\L'})  \Shell_{OO'} [\dlin_\L] (\vx_1,\vx_2)   \,, \nonumber
 \ea
where we see that terms $\Shell_{OO'}$ appear. Finally, the $m=3$ correction is
\ba
& \Zcorr^{(3)}[J_\L, \dlin_\L] =  \frac16 \int_{\vx}  \left[ J_{\L}(\vx) \right]^3   \sum_{O} B_{\eps,O}({\L'})  \Shell_{O} [\dlin_\L](\vx)   \vs
&\hspace{2cm} + \frac12 \int_{\vx_1,\vx_2}  \left[ J_{\L}(\vx_1) \right]^2 J_{\L}(\vx_2)    \sum_{O, O'} b_O({\L'}) P_{\eps,O'}({\L'})  \Shell_{OO'} [\dlin_\L] (\vx_1,\vx_2)    \label{eq:corr3} \\
&\hspace{2cm} + \frac16 \int_{\vx_1,\vx_2,\vx_3}  J_{\L}(\vx_1) J_{\L}(\vx_2) J_{\L}(\vx_3)  \vs 
& \hspace{4cm}\sum_{O, O', O''} b_O({\L'}) b_{O'}({\L'})b_{O''}({\L'})  \Shell_{OO'O''} [\dlin_\L] (\vx_1,\vx_2,\vx_3)     \,. \nonumber
\ea
We highlight again that terms proportional to $J^1$ (i.e., the bias operators) in general source $J^2,J^3,$ etc. operators. The terms proportional to $J^2$ (the Gaussian stochastic terms $P_{\eps,O}$) source $J^3,J^4$, etc. terms. Therefore, each order in $J^m$ is only affected by powers of $J$ smaller than $m$, as already pointed out by \cite{Carroll:2013oxa} for the matter case. Later, this will allow us to obtain a self-consistent solution of the RG evolution.

The single, double and triple-operator shell contraction $\Shell_{O}$, $\Shell_{OO'}$  and $\Shell_{OO'O''}$ are defined in Fourier space as
\ba
\Shell_{O}(\vk) &= \sum_{\substack{i\geq2 \\ i \,\textrm{even}}} \Shell_{O}^{i}(\vk) =  \Shell_{O}^{2} [\dlin_\L](\vk) + \Shell_{O}^{4} [\dlin_\L](\vk) + \dots\,,
\\
\Shell_{OO'}(\vk,\vk') &= \sum_{\substack{i\geq1,i'\geq1 \\ i+i' \,\textrm{even}}} \Shell_{OO'}^{ii'}(\vk,\vk')  = \Shell_{OO'}^{11} [\dlin_\L] (\vk,\vk')  \\
& \hspace{4cm }+ 2\Shell_{OO'}^{13} [\dlin_\L] (\vk,\vk') + \Shell_{OO'}^{22} [\dlin_\L] (\vk,\vk') + \dots\,,
\vs
\Shell_{OO'O''}(\vk,\vk',\vk'') &= \sum_{\substack{i\geq1,i'\geq1,i''\geq1 \\ i+i'+i'' \,\textrm{even}}} \Shell_{OO'O''}^{ii'i''}(\vk,\vk',\vk'') = 3 \Shell_{OO'O''}^{112} [\dlin_\L] (\vk,\vk',\vk'')  \\
& \hspace{2cm }+ \Shell_{OO'O''}^{222} [\dlin_\L] (\vk,\vk',\vk'') + 6\Shell_{OO'O''}^{123} [\dlin_\L] (\vk,\vk',\vk'') + \dots \,.\nonumber
\ea
The corresponding diagrams are shown in \reffigs{SOdiag}{SOOOdiag} below.
Notice that we keep only the terms involving even powers of shell fields, given our assumption of Gaussian initial conditions, with
\ba
\Shell_{O_1\dots O_\o}^{i_1 \dots i_\o} [\dlin_\L] (\vk_1,\dots, \vk_{\o}) &= \label{eq:shellgen}\\ 
 \sum_{n_1\geq i_1,\dots, n_\o \geq i_\o} & \left\< O_1^{(n_1),(i_1)_{\rm shell}} [\dlin_\L,\dlinshell](\vk_1) \ldots O_\o^{(n_\o),(i_\o)_{\rm shell}} [\dlin_\L,\dlinshell](\vk_\o) \right\>_{\rm shell} \,. \nonumber
 \ea

\paragraph{General structure of partition function and its renormalization.} 
Let us consider the part of the effective action \refeq{Sreal} that involves precisely $m$ powers of the current, and compute how the correction $\Zcorr^{(m)}$ renormalizes the coefficients. We have:
\ba
 S[J_{\L}] & \supset \frac{1}{m!} \int_{\vx} \left[ J_{\L}(\vx) \right]^m   \sum_{O} C^{(m)}_O({\L'})  O[\dlin_\L](\vx) + \Zcorr^{(m)}[J_\L, \dlin_\L]\,, \label{eq:generalS}
\ea
where, again, $C^{(m)}_O({\L'})$ is the stochastic coefficient of $m$-th order in $J$, related to the $m$-th cumulant of the stochasticity. 
The correction at $m$-th order in currents can be written as a sum in the number of operators $\o$ that are being contracted, where for each $\o$ we sum over the contributions $\Shell_{O_1 \dots O_\o}$:
{
\begin{equation}
\boxed{
\begin{aligned}
& \Zcorr^{(m)}[J_\L, \dlin_\L] =  \frac{1}{m!}  \sum_{\o=1}^m \sum_{\substack{i_1=1,\cdots, i_p=1 \\ i_1+\dots+ i_p = m}}^m \frac{\prod_{a=1}^m\mathcal{N}(a,\{i_1,\dots,i_{p}\})!}{p!} g^{i_1,\dots,i_{p}}_m
 \\
&\hspace{.5cm}   \int_{\vx_{1},\dots,\vx_{p}} \left[J_{\L}(\vx_1)\right]^{i_1}\dots \left[J_{\L}(\vx_\o)\right]^{i_\o}  \sum_{O_1,\dots,O_\o} C^{(i_1)}_{O_1}({\L'})\dots C^{(i_{\o})}_{O_{\o}}({\L'}) \Shell_{O_1 \dots O_\o}^{2\dots 2} [\dlin_\L] \left(\vx_1,\dots,\vx_\o\right)  \,, \label{eq:generalcorr}
\end{aligned}
} 
\end{equation}}
where
\ba
g^{i_1,\dots,i_{p}}_m \equiv  m! \prod_{a=1}^m \left[\frac{1}{\mathcal{N}(a,\{i_1,\dots,i_{p}\})!}\right] \left[\frac{1}{a!}\right]^{\mathcal{N}(a,\{i_1,\dots,i_{p}\})}\,,  \label{eq:symmetrydef}
\ea
is a symmetry factor in which $\mathcal{N}(a,\{i_1,\dots,i_{\o}\})$ is the number of times $a$ appears in $\{i_1,\dots,i_{\o}\}$; hence, $g^{1}_1 = 1$, $g^{11}_2 = 1$, $g^{2}_2 = 1$, $g^{111}_3 = 1$, $g^{12}_3 = g^{21}_3 = 3$, $g^{3}_3 = 1$, and so forth. 
Notice that we isolated a factor $\frac{1}{m!}\prod_{a=1}^m \left[\frac{1}{\mathcal{N}(a,\{i_1,\dots,i_{p}\})!}\right]$ in the first line of \refeq{generalcorr} for reasons that will become clear in \refsec{summary}.
\refeq{generalcorr} is the master equation that will guide the following sections. 
One can trivially recover \refeq{corr1}, \refeq{corr2} and \refeq{corr3} for $m=1,2,3$ respectively. On the other hand, it is important to stress the difference with respect to Eq.~(4.22) of \cite{Carroll:2013oxa}, which only includes the contributions of the type $S^{2}_{O}$ and $S^{11}_{OO'}$.
As discussed in \refsec{introsummary}, our goal here is to take the external momenta in the current to zero, to obtain a partition function that only has local interactions. In that case, as we will see later and already anticipated in \refeq{generalcorr}, we need to include diagrams with more than one shell propagator, but \emph{only} the specific contributions $\Shell_{O_1 \dots O_\o}^{2\dots 2}$. In the same limit of low external momenta, the $S^{11}_{OO'}$ term is instead suppressed due to kinematics and the orthogonality with the current $J_\L$.

\subsection{On the \texorpdfstring{$\Shell_{O'}$}{SO} corrections to \texorpdfstring{$O$}{O} [\texorpdfstring{$p=1$}{p1} in \refeq{generalcorr}]} \label{sec:SO}

\begin{figure}[t]
    \centering
    \includegraphics[width = 0.2\textwidth]{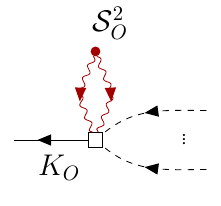}\hspace*{2cm}
    \includegraphics[width = 0.27\textwidth]{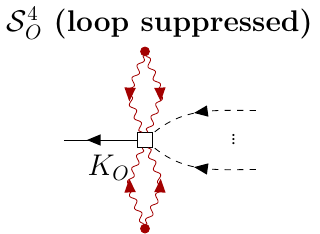}
    \caption{Diagrams for the $\Shell_O^2$ (1-loop) and $\Shell_O^4$ (2-loop) contributions.}
    \label{fig:SOdiag}
\end{figure}

The contributions from $\Shell_{O'}$ enter generically in the correction of all coefficients $\Zcorr^{(m)}$ defined in \refeq{generalcorr} when taking $\o = 1$. As example, we calculate the correction to the $J^1$ coefficients $b_O = C_O^{(1)}$, but that contribution in fact appears for any coefficient $C_{O}^{(m)}$. 
We display the first $\Shell_{O}^{i}$ diagrams in \reffig{SOdiag}, in which we see the 1-loop $\Shell_{O}^{2}$ diagram (left) and the $\Shell_{O}^{4}$ diagram (right), the last being suppressed as described in \refapp{S4suppression}.  

\paragraph{The $\Shell_{O'}^2$ terms.}
We begin with the most general structure of RG contributions coming from $\Shell_{O'}^2[\dlin_\L]$. This part follows the derivations of Paper 1, but we reproduce it for completeness. In general, we can write for the $\Shell_{O'}^2[\dlin_\L]$ operator
\be
\Shell_{O'}^2[\dlin_\L] = \leps  \frac{d\s^2_\L}{d\L}\Big|_{\L} \sum_{O} s_{O'}^O O[\dlin_\L],
\label{eq:SO2exp}
\ee
where $s_{O'}^O$ is the contribution of the operator $O'$ to $O$ via $\Shell_{O'}^2$.
We used the results of Paper~1, reproduced in \refapp{shell}, in which we find that the single-operator shell integrals $\Shell_{O'}^2$ are proportional to $\int_{\vp} P_{\rm shell}(p)$ after expanding
\be
\int_{\vp} P_{\rm shell}(p) = \int_{\L}^{\L+\leps} \frac{p^2 dp}{2 \pi^2 }\Plin(p) = \frac{d \s^2_\L}{d \L}\Big|_{\L}\leps + \O(\leps^2)\,.
\label{eq:sigmashell}
\ee
Therefore, the correction to $b_O (\L) {O}[\dlin_{\L}]$ from $\Shell_{O'}^2[\dlin_\L]$ in \refeq{generalS} is given by 
\be
\sum_O b_O(\L) O[\dlin_\L] = \sum_O \left[ b_O({\L'}) + \leps \frac{d \s^2_\L}{d \L}\Big|_{\L} \sum_{O'} s_{O'}^O  b_{O'}(\L')\right] O[\dlin_\L] \,.
\ee
We can then identify 
\be \label{eq:b_difftemplate}
b_O(\L) = b_O(\L') + \leps \frac{d \s^2_\L}{d \L}\Big|_{\L} \sum_{O'} s_{O'}^O  b_{O'}(\L')\, ,
\ee
and after writing $\L' = \L + \leps$ and taking $\leps \to 0$, we can write
\be
\boxed{
\frac{d}{d\L} b_O(\L) = - \frac{d\s^2_\L}{d\L} \sum_{O'} s_{O'}^O \, b_{O'}(\L)\, , \label{eq:S2forallorder}
}
\ee
where the derivative of the variance is evaluated at $\L$.
The set of coefficients $s_{O'}^O$ is presented in \reftab{cCoeff}, as described in \refapp{S2eval} and Paper 1. 

\begin{table}[ht]
	\centering
	{ \begin{tabular}{|c||c|c|c|c|c|c|c|}
			\hline
			$s_{O'}^O$ & $\d$ & $\d^2$  & $\G_2$ & $\d^3$ & $\G_3$ & $\Gamma_3$ & $\d\G_2$  \\
			\hline\hline
			$\unit$ &   -& {-}& -& -& -& -& -\\
			\hline
			$\d$ &   -& $ 68/21$&-& 3& -& -& $-4/3$\\
			\hline
			$\d^2$ &  - & $ 8126/2205$&- & $68/7$&- &- &$- 376/105$ \\
			\hline
			$\G_2$ &  - & $254/2205$&- &- &- &- &$ 116/105$ \\
			\hline
	\end{tabular}}
	\caption{The coefficients $s_{O'}^O$ summarizing the single-operator shell contractions $\Shell_{O'}^2$ to the operator $O$, with $O'$ in the columns and $O$ in the lines.  }
	\label{tab:cCoeff}
\end{table}

A last interesting point to notice here is that the corrections of the type $\Shell_{O'}$ [$\o=1$ in \refeq{generalcorr}] do not generate contributions to the zeroth-order operator $C_{\unit}^{(m)}$, i.e. the tadpole, which reflects the fact that $\langle O \rangle = 0$ for all operators. The terms $C_{\unit}^{(m)}$ are therefore only sourced by higher-order contributions starting from $\Shell_{O'O''}$.

\paragraph{The (suppression of) $\Shell_{O'}^4$ and higher-order terms.}
We dedicate \refapp{S4suppression} to discuss how the contributions coming from a single operator but expanded in higher orders of shell fields  (2- and higher-loop terms: $\Shell_{O'}^4$, $\Shell_{O'}^6$, $\dots$) are suppressed by extra factors of $\lambda$ compared to $\Shell_{O'}^2$. Since we can take the limit $\leps\to 0$ without loss of generality of the RG flow, we can therefore drop these terms, and it is sufficient to consider $\Shell_{O'}^2$ in $\Zcorr^{(m)}[J_\L, \dlin_\L]$.

\subsection{On the \texorpdfstring{$\Shell_{O'O''}$}{SOO} corrections to \texorpdfstring{$O$}{O} [\texorpdfstring{$p=2$}{p2} in \refeq{generalcorr}]} \label{sec:SOOmain}

\begin{figure}[t]
    \centering
    \includegraphics[width = 0.4\textwidth]{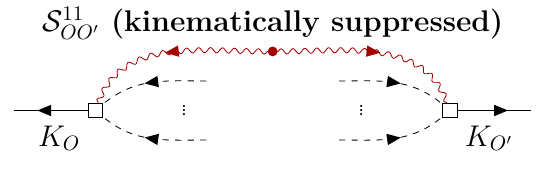}
    \includegraphics[width = 0.4\textwidth]{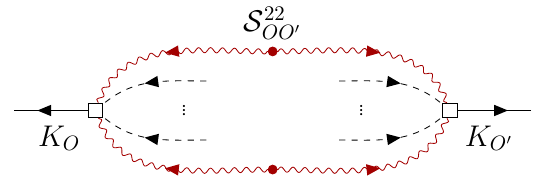}
    \includegraphics[width = 0.4\textwidth]{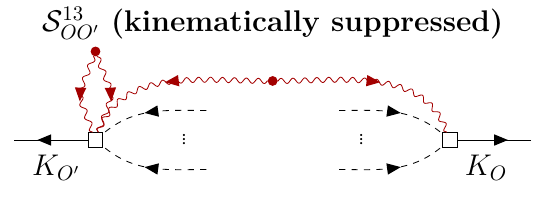}
    \caption{Diagrams for the $\Shell_{OO'}$ terms. We display the tree-level contribution $\Shell_{OO'}^{11}$ (top left) and the two 1-loop contributions $\Shell_{OO'}^{22}$ (top right) and $\Shell_{OO'}^{13}$ (bottom).}
    \label{fig:SOOdiag}
\end{figure}

The double-operator shell contribution $\Shell_{O'O''}$ acts as a source to $\O(J^2)$ terms in the action, as well as higher-order terms, as one can see from \refeq{corr2}. 
We consider the coefficients $P_{\eps,O} = C^{(2)}_{O}$ here as example.
To keep the structure clear, in this section we keep only $\Shell_{O'O''}$ as a source of $P_{\eps,O}$. 
The tree-level and 1-loop diagrams for the $\Shell_{OO'}^{ii'}$ terms are shown in \reffig{SOOdiag}.

\paragraph{The main contributions from $\Shell_{O'O''}$.} 

We start by discussing which of the diagrams described in \reffig{SOOdiag} constitute the main stochastic contribution. 
In order to not overload the main text with the technical details, we dedicate \refapp{doubleshell} to show that both the tree-level term $\Shell_{O'O''}^{11}$ and the 1-loop term $\Shell_{O'O''}^{13}$ are zero due to their kinematic structure.
This can already be seen from the top left and bottom diagrams in \reffig{SOOdiag}: all dashed incoming lines to the $K_O$ involve $\dlin_\L$, and are thus constrained to have momenta $\leq \L$. On the other hand, the outgoing lines are constrained by the support of the current factors $J$, i.e. by the external momenta $\vk,\vk'$. In the limit of $k,k' \ll \L$, there is a vanishing amount of phase space available for momenta in the shell (wiggly lines) in this kinematic configuration.
Hence, these contributions cannot form the dominant source of stochastic terms.

\begin{figure}[t]
    \centering
    \includegraphics[width = 0.32\textwidth]{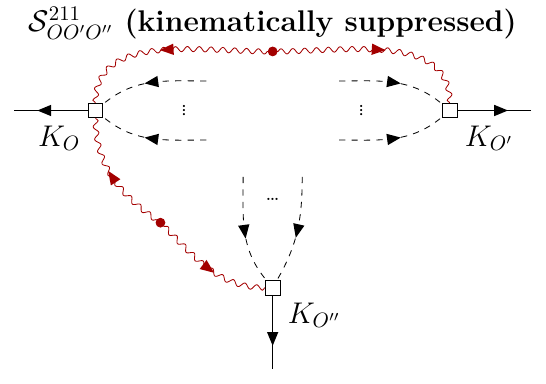}
    \includegraphics[width = 0.32\textwidth]{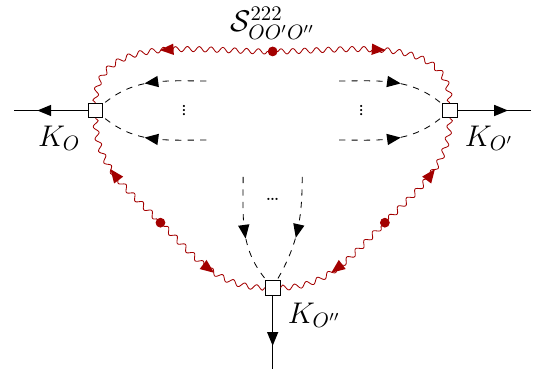}
    \includegraphics[width = 0.32\textwidth]{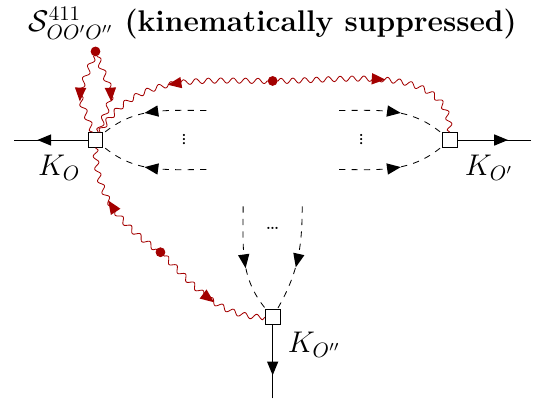} 
    \includegraphics[width = 0.32\textwidth]{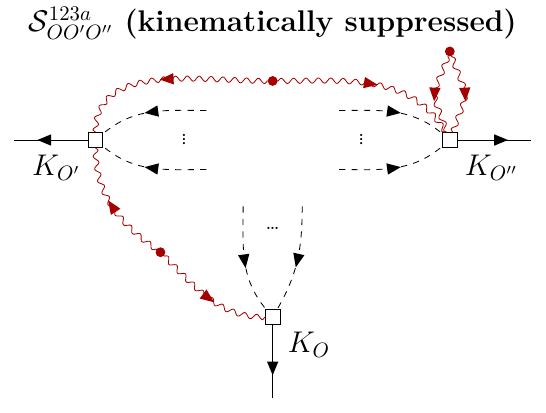} 
    \includegraphics[width = 0.32\textwidth]{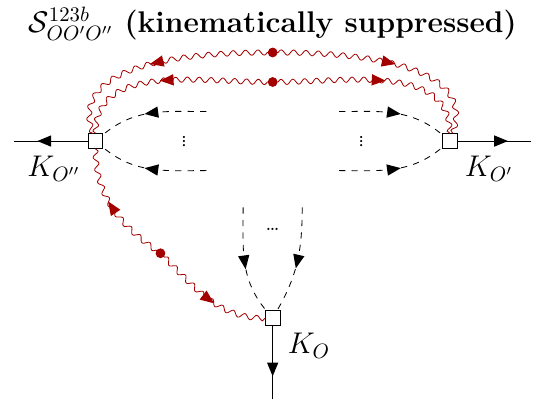} 
    \caption{Diagrams for the $\Shell_{OO'O''}$ terms. The tree-level contribution $\Shell_{OO'O''}^{111}$ is displayed in the top-left diagram. The other four diagrams represent the 1-loop contributions $\Shell_{OO'O'''}^{222}$, $\Shell_{OO'O'''}^{114}$ and $\Shell_{OO'O'''}^{123}$. }
    \label{fig:SOOOdiag}
\end{figure}

In the limit of soft external moments, the leading contribution instead comes from the $\Shell_{O'O''}^{22}$ term.
This was in fact already anticipated in Eq.~(3.13) of \cite{Cabass:2019lqx}.
As is apparent from the the top right diagram in \reffig{SOOdiag}, the
  kinematic suppression does not apply in this case, as two modes with momenta in the shell can couple to produce an external momentum $k\ll\L$.
In fact, this is a specific case of the generic conclusion of \refapp{mostgeneralstruct} that the terms of the type $\Shell_{O_1O_2 \dots O_\o}^{22\dots2}$ are the leading sources in general.
To exemplify this, we show the corresponding tree-level and 1-loop diagrams for the $\Shell_{OO'O''}$ terms in \reffig{SOOOdiag}.

\paragraph{Corrections to $P_{\eps,\unit}$ and $P_{\eps,\d}$.}

We now focus on contributions to zeroth and first-order leading-in-derivative ($d_{O} = 0$) operators $P_{\eps,\unit}$ and $P_{\eps,\d}$.
Since these contributions will exemplify the general structure of RG source terms, we go into a bit more technical detail here. Readers only interested in the final result can skip ahead to \refeq{generalP}.

In this work we focus only on contributions sourced by up to second-order operators.  We show in \refapp{SOOcalc} that, when neglecting operators starting from third-order the only of the $\Shell_{O'O''}^{22}$ diagrams that contributes to those operators is $\Shell_{\d^2\d^2}^{22}$ via
\ba
\Zcorr^{(2)} &\supset [b_{\d^2}(\L)]^2\int_{\vk,\vk'} J_\L(\vk) J_\L(\vk') \Shell^{22}_{\d^2\d^2}(\vk,\vk') \,.
\ea
Using the result of \refeq{S22_d2d2_calc}, we find that the leading contribution of $\Shell^{22}_{\d^2\d^2}$ to the zeroth-order operator is 
\ba
\Shell^{22}_{\d^2\d^2}(\vk,\vk') &\supset 2 \diracpi(\vk+\vk') \int_{\vp} P_{\rm shell}(p) P_{\rm shell}(|\vk-\vp|) ,
\ea
with corrections of order $(k/\L)^2$, which are absorbed by higher-derivative stochastic contributions. 
Despite the fact that this correction involves four powers of $\dlinshell$, it is linear order in $\lambda$, i.e. of the same order as $\Shell_O^2$, since
\ba 
\int_{\vp} P_{\rm shell}(p) P_{\rm shell}(|\vk-\vp|) \stackrel{k/p \to 0}{\approx} \int_{\vp} [P_{\rm shell}(p)]^2 = {2} \Plin(\L) \frac{d\s^2_\L}{d\L} \lambda \,.
 \label{eq:linearinlambda}
\ea
Following \refeq{corr2}, this yields a correction to the term
\ba 
P_{\eps,\unit} \int_{\vk} J_\L(\vk)J_\L(-\vk) \,,
\ea
which reads
\ba 
\Zcorr^{(2)}[J_\L, \dlin_\L]  &\supset \frac12 2 \left[b_{\d^2} \right]^2 \int_{\vk} J_\L(\vk)J_\L(-\vk)  \int_{\vp} P_{\rm shell}(p) P_{\rm shell}(|\vk-\vp|) \vs
&\approx \frac12 2 [b_{\d^2}(\L)]^2  \int_{\vp} [P_{\rm shell}(p)]^2 \int_{\vk} J_\L(\vk)J_\L(-\vk) \vs 
&=  \frac12 s_{\d^2\d^2}^\unit [b_{\d^2}(\L)]^2  \left[ {2} \Plin(\L) \right] \frac{d\s^2_\L}{d\L} \lambda \int_{\vk} J_\L(\vk)J_\L(-\vk) \,,
\ea
where in the second line we used $p \gg k$ and we introduced the shell contribution $s_{\d^2\d^2}^\unit = 2$, which determines the contribution of $\Shell^{22}_{\d^2\d^2}$ to $P_{\eps,\unit} J^2\unit$. 
Moreover, it scales as expected from dimensional analysis (recall that $P_{\eps,O}$ has dimensions of power spectrum, i.e. length cubed).
Therefore, following the same steps as \refeq{b_difftemplate}, we find 
\be
  \boxed{
    \frac{d}{d\L} P_{\eps,\unit}(\L) = - s_{\d^2\d^2}^\unit [b_{\d^2}(\L)]^2 \left[ {2} \Plin(\L) \right]   \frac{d\s^2_\L}{d\L} \, .
  }
    \label{eq:diffPunit}
\ee

We now move to the corrections to the first-order operator $P_{\eps,\d}$. 
As calculated in \refeq{S22d2d2oneleg}, we find that the contribution of $\Shell^{22}_{\d^2\d^2}$ to the term $P_{\eps,\d}J^2 \d$ in the effective action is 
\ba
\Shell^{22}_{\d^2\d^2}(\vk,\vk') &\supset 8 \dlin_\L(\vk+\vk') \int_{\vp}F_2(\vk+\vk', -\vp)P_{\rm shell}(p)P_{\rm shell}(|\vk - \vp|)   \vs
&\stackrel{k/p \to 0}{\approx} s_{\d^2\d^2}^\d\dlin_\L(\vk+\vk')  \left[ {2} \Plin(\L) \right] \frac{d\s^2_\L}{d\L} \lambda \,,
\ea
after using \refeq{lininlambda} and defining $s_{\d^2\d^2}^\d = 136/21$, that determines the contribution of $\Shell^{22}_{\d^2\d^2}$ to $\d$. Similarly, using \refeq{S22d3d2oneleg} and \refeq{S22G2d2oneleg}
\ba
\Shell^{22}_{\d^3\d^2}(\vk,\vk') &\supset 6 \dlin_\L(\vk+\vk') \int_{\vp} P_{\rm shell}(p)P_{\rm shell}(|\vk - \vp|)   \vs
&\stackrel{k/p \to 0}{\approx} s_{\d^3\d^2}^\d\dlin_\L(\vk+\vk')  \left[ {2} \Plin(\L) \right] \frac{d\s^2_\L}{d\L} \lambda \,, \\
\Shell^{22}_{(\G_2\d)\d^2}(\vk,\vk') &\supset 4 \dlin_\L(\vk+\vk') \int_{\vp} \s^2_{-\vp,\vk+\vk'} P_{\rm shell}(p)P_{\rm shell}(|\vk - \vp|)   \vs
&\stackrel{k/p \to 0}{\approx} s_{(\G_2\d)\d^2}^\d\dlin_\L(\vk+\vk')  \left[ {2} \Plin(\L) \right] \frac{d\s^2_\L}{d\L} \lambda \,,
\ea
with $s_{\d^3\d^2}^\d = 6$ and $s_{(\G_2\d)\d^2}^\d = -8/3$. 
Following \refeq{corr2}, those terms yield a correction to
\ba 
P_{\eps,\d} \int_{\vk, \vk'} J_\L(\vk)J_\L(\vk') \dlin_\L(\vk+\vk')  \,,
\ea
that reads
\ba 
\Zcorr^{(2)}[J_\L, \dlin_\L]  &\supset \frac12 \left[ s_{\d^2\d^2}^\d b_{\d^2}(\L)b_{\d^2}(\L) + s_{\d^3\d^2}^\d b_{\d^3}(\L)b_{\d^2}(\L) + s_{(\G_2\d)\d^2}^\d b_{\G_2\d}(\L)b_{\d^2}(\L)\right]\vs 
& \hspace{3cm} \times  \left[ {2} \Plin(\L) \right] \frac{d\s^2_\L}{d\L} \lambda \int_{\vk,\vk'} J_\L(\vk)J_\L(\vk') \dlin(\vk+\vk') \,.
\ea
Therefore, considering only the $\Shell_{O'O''}$ contributions we find
\ba 
&\frac{d}{d\L} P_{\eps,\d}(\L) = -  \left[ {2} \Plin(\L) \right]  \frac{d\s^2_\L}{d\L} \left[ s_{\d^2\d^2}^\d b_{\d^2}(\L)b_{\d^2}(\L) + s_{\d^3\d^2}^\d b_{\d^3}(\L)b_{\d^2}(\L) + s_{(\G_2\d)\d^2}^\d b_{\G_2\d}(\L)b_{\d^2}(\L)\right] \, .  \label{eq:diffPd} 
\ea

In conclusion, similarly to \refeq{diffPunit} and \refeq{diffPd}, we can then write for the general case
\be 
\boxed{
\frac{d}{d\L} P_{\eps,O}(\L) = -  \left[ {2} \Plin(\L) \right]  \frac{d\s^2_\L}{d\L}  \sum_{O',O''}s_{O'O''}^O [b_{O'}(\L)][b_{O''}(\L)] \label{eq:generalP}  \,, 
}
\ee
where $s_{O'O''}^O$ is the contribution of the operator $O'$ and $O''$ to $O$ via $\Shell_{O'O''}^{22}$. 
The $s_{O_1O_2}$ coefficients that contribute to zeroth and first-order are
\ba
s_{\d^2\d^2}^\unit = 2 \,, \quad
s_{\d^2\d^2}^\d = \frac{136}{21} \,, \quad 
s_{\d^3\d^2}^\d = 6 \,, \quad 
s_{(\G_2\d)\d^2}^\d = -8/3 \,. \label{eq:sOOvalues}
\ea

\subsection{The general source structure from the master equation [\refeq{generalcorr}]} \label{sec:summary}

One can easily show using \refeq{generalcorr} that we can construct the general system of ODEs for all the bias and stochastic parameters to be
\begin{equation}
\boxed{
\begin{aligned}
& \frac{d}{d\L} C_{O}^{(m)}(\L) =     \sum_{\o=1}^m \sum_{\substack{i_1=1,\cdots, i_p=1 \\ i_1+\dots+ i_p = m}}^m \frac{\prod_{a=1}^m\mathcal{N}(a,\{i_1,\dots,i_{p}\})!}{p!}  {\o} \left[\Plin(\L)\right]^{\o-1}\frac{d\s^2_\L}{d\L}   \label{eq:masterSol} \\
&\hspace{3cm} \times g^{i_1,\dots,i_{p}}_m \sum_{O_1,\dots,O_\o} s_{O_1O_2 \dots O_\o}^O  C^{(i_1)}_{O_1}({\L})\dots C^{(i_{\o})}_{O_{\o}}({\L})   \,,
\end{aligned}
}
\end{equation}
where $s_{O_1O_2 \dots O_m}^O$ is the contribution of the operators $O_1\,,O_2\,, \dots\,,O_m$ to $O$ via $\Shell_{O_1 O_2 \dots O_m}^{22\dots2}$. In addition to the $s_{O_1}$ and $s_{O_1O_2}$, described in \reftab{cCoeff} and \refeq{sOOvalues}, we now have
\ba
s_{\d^2\d^2\d^2}^\unit = 8 \,, \quad
s_{\d^2\d^2\d^2}^\d = \frac{544}{21} \,, \quad
s_{\d^3\d^2\d^2}^\d = 24\,, \quad
s_{(\G_2\d)\d^2\d^2}^\d = -\frac{32}{3} \,,
\ea
as calculated in \refapp{SOOOeval}. 
Notice that the prefactor $\frac{\prod_{a=1}^m\mathcal{N}(a,\{i_1,\dots,i_{p}\})!}{p!}$ is simply a permutation factor and only $g^{i_1,\dots,i_{p}}_m$ will appear in the final ODEs. 
Another key fact is that $\Shell_{O_1 O_2 \dots O_m}^{22\dots2}$ terms contribute all at the same order (i.e. linearly in $\lambda$) as a consequence of 
\be 
\boxed{
\int_{\vp} P_{\rm shell}(p) P_{\rm shell}(|\vk_1-\vp|) \dots P_{\rm shell}(|\vk_{1\dots i}-\vp|) \stackrel{k/p \to 0}{\approx}   {(i+1)} \left[ \Plin(\L)\right]^{i} \frac{d\s^2_\L}{d\L} \lambda  \,, \label{eq:lininlambda}
}
\ee
that generalizes \refeq{linearinlambda}, such that
\bea 
\frac{d}{d\L} C_{O}^{(m)}(\L) &\propto& - \left[ \Plin(\L)\right]^{p-1}\frac{d\s^2_\L}{d\L} \sum_{O_1,O_2,\dots O_m} s_{O_1O_2 \dots O_m}^O   \, C^{(i_1)}_{O_1}({\L})\dots C^{(i_{\o})}_{O_{\o}}({\L})\,, \vs
&& \hspace{7cm} \textrm{via} \quad \Shell_{O_1 O_2 \dots O_\o}^{22\dots2}\,.
\eea
Notice the specific scaling of each contribution with different powers of $\left[ \Plin(\L)\right]^{\o-1}$, which is required by dimensional analysis [\refeq{dimC}]. 
We emphasize once more the importance of \refeq{masterSol}: {\it it governs the general RG flow of all leading-in-derivatives EFT parameters}. It makes manifest how any coefficient (including bias and stochastic parameters) of the EFT of LSS is sourced by a variation in the smoothing scale $\L$. Whether the RG running can lead to more (non-perturbative) information, such as in the context of quantum field theory, will be explored in a further publication. We refer to Sec.~4 of Paper 1 for a comparison to other renormalization schemes.  

Let us consider the zeroth-order stochastic contributions $C_\unit^{(m)}$, which correspond to the moments of the effective stochastic field [\refeq{Cmunit}].
We find that the only contribution comes from $\Shell_{\d^2\d^2 \dots \d^2}^{22\dots2}$ [noticing that the contribution $\Shell^{2} = 0$ after removing the tadpole using \refeq{Onorm}], leading to 
\begin{equation}
\boxed{
\begin{aligned}
& \frac{d}{d\L} C_{\unit}^{(m)}(\L) =   \sum_{\o=2}^m \sum_{\substack{i_1=1,\cdots, i_p=1 \\ i_1+\dots+ i_p = m}}^m \frac{\prod_{a=1}^m\mathcal{N}(a,\{i_1,\dots,i_{p}\})!}{p!} g^{i_1,\dots,i_{p}}_m   \\
& \hspace{3cm} \times  {p} \left[ \Plin(\L)\right]^{p-1}\frac{d\s^2_\L}{d\L} s_{\d^2\d^2 \dots \d^2}^\unit C^{(i_1)}_{\d^2}({\L})\dots C^{(i_{\o})}_{\d^2}({\L}) \label{eq:dUnit} \,. 
\end{aligned}
}
\end{equation}
This leads to an important conclusion: \emph{the presence of the operator $\d^2$
in the bias expansion immediately generates all $C_\unit^{m>1}$, i.e. all moments of the stochasticity.} 

We display the structure of the RG source terms in \reffig{sourcestructure}. The columns show different operator orders, while the rows the different powers in $J$. The arrows $A\to B$ indicate which terms $B$ in the action are sourced by a given term $A$.

\subsection{Summary} 

\begin{figure}[t]
	\centering
	\includegraphics[width = 0.6\textwidth]{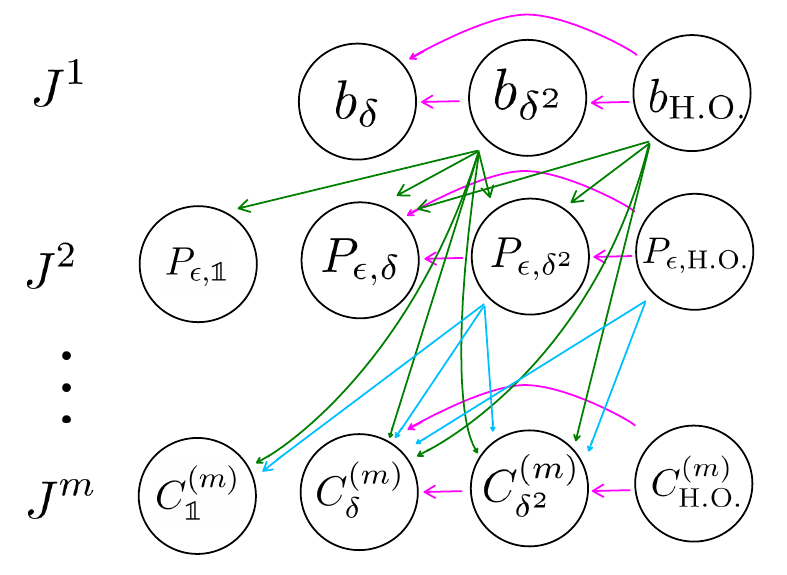}
	\caption{Diagram showing connections between parameters at different orders in $J$ (rows), and different operator orders (columns).  Arrows point from the source to the sourced operator. The last circle in each line indicates higher-order (H.O.) operators. The magenta arrows indicate contributions from $\Shell_O^2$ (i.e. within the same order $m$), with the other colors representing other types of contributions via $\Shell_{O_1 \dots O_p}^{2\dots 2}$. Only leading-in-derivative terms are shown in the figure. }
	\label{fig:sourcestructure}
\end{figure}

We now summarize the main conclusions of this section, providing some insights into the structure of the general RG-LSS equation:
\begin{itemize}
\item We provide a master formula \refeq{generalcorr} which describes the source term of the RG equations of stochastic terms \emph{at all orders}, i.e. at all powers $m$ in the current $J$. This leads to the complete set of RG equations described by \refeq{masterSol}.
\item In general, $m$-th order stochastic moments contribute to $n$-th order moments with $m \geq n$. For example, bias terms ($m=1$) lead to runnning of bias coefficients as well as (in general) all stochastic contributions.
For the specific case of $n=m$, the $\Shell^2_O$ terms are the only contribution. Both of these points were already noticed by \cite{Carroll:2013oxa}.
  For $n=m=1$, we recover the results of Paper 1, which are represented by the first row in \reffig{sourcestructure}.
\item Zeroth and first-order operators do not contribute to leading-in-derivative operators at any order in $J$. They only source higher-derivative operators (not shown in \reffig{sourcestructure}). Notice that no arrows leave those operators in \reffig{sourcestructure}.
\item If all source terms to the RG equations were of the $\Shell^2_O$ type, there would be no coupling between terms with different powers in the current, $J^m$. In particular, no stochasticity ($m\geq 2$) would be generated by nonlinear bias ($m=1$). However, we find that terms of the form $\Shell_{O_1 \dots O_m}^{2\dots2}$ generate contributions of similar relevance as $\Shell^2_O$ to all higher $m$.
  Thus, according to our RGE system, a single nonlinear bias term $b_{\delta^2}(\Lambda_\ast)$ immediately generates \emph{all stochastic moments} when running to a scale $\Lambda < \Lambda_\ast$.
  In contrast, Ref.~\cite{Carroll:2013oxa} found that higher-order stochastic terms are generated by successive ``rounds'' of the RG flow, as they only considered $\Shell_{O}^{2}$ and $\Shell_{O_1 O_2 }^{11}$ contributions. As discussed in \refsec{introsummary}, the difference arises because the intermediate state in the approach of \cite{Carroll:2013oxa} involves \emph{nonlocal interactions}, while our RGE are written directly in terms of the local interaction terms that emerge in the IR limit.
\end{itemize}
In short, we have derived the most general running of the EFT of LSS coefficients in \refeq{masterSol}. 
This result can be straightforwardly used (both at field-level and by $n$-point functions) to validate the running of the freee EFT coefficients. It can also be used to renormalize any $n$-point function at any loop order including stochastic terms.
Our result goes beyond the usual $n$-point function renormalization considered by other works considering the full RG running of the parameters, with the $n$-point function normalization being the $\L \to 0$ limit of the RG running (see Sections~3 and 4 of \cite{Rubira:2023vzw} for a connection between both schemes).   

\section{Results}
\label{sec:results}

After having provided the general set of equations for the running of bias and stochastic parameters in \refeq{masterSol}, we move to discuss their solutions. We separate the discussion into the different powers of $J$.

\paragraph{$\bm{J^1}$: bias parameters.}
For the case $J^1$, we find that only the $s_{O'}^O$ in \reftab{cCoeff} contribute, such that, following Paper 1
\ba
\frac{d b_\d}{d \L} &= -\left[\frac{68}{21}b_{\d^2}(\L)+b_{n=3}^{\ast\{\d\}}\right]\frac{d \s^2_\L}{d \L} + \O\left[ \partial_\L b_{n=3}^{\{\d\}}\right]\,, \label{eq:drun}
\\
\frac{d b_{\d^2}}{d \L} &= -\left[\frac{8126}{2205}b_{\d^2}(\L) + b_{n=3+4}^{\ast\{\d^2\}}  \right]\frac{d \s^2_\L}{d \L} + \O\left[ \partial_\L b_{n=3+4}^{\{\d^2\}}\right]\,, \label{eq:d2run}
\\
\frac{d b_{\G_2}}{d \L} &= -\left[  \frac{254}{2205 }b_{\d^2}(\L) + b_{n=3+4}^{\ast\{\G_2\}} \right]\frac{d \s^2_\L}{d \L} + \O\left[ \partial_\L b_{n=3+4}^{\{\G_2\}}\right]\,. \label{eq:Grun}
\ea
We use the notation $b^\ast = b(\L_\ast)$ for the parameters that are evaluated at a fixed renormalization scale $\L_\ast$ and\footnote{Notice that we switched the parenthesis in Paper 1 for curly brackets in terms such as $b_{n=3+4}^{\{\d\}}$ to avoid confusion with the $C^{(m)}$ coefficients.}
\ba
b_{n=3+4}^{\{\d\}} &= 3b_{\d^3}-\frac{4}{3}b_{\G_2\d} \,, \label{eq:hobd} \\
b_{n=3+4}^{\{\d^2\}} &= \frac{{68}}{7}b_{\d^3}  - \frac{376}{105}b_{\G_2\d} + b_{n=4}^{\{\d^2\}}  \,, \\
b_{n=3+4}^{\{\G_2\}} &= \frac{116}{105}b_{\G_2\d} + b_{n=4}^{\{\G_2\}} \label{eq:hobG2}\,,
\ea
to account for contributions from higher-order operators in a short-hand notation, in which $b_{n=n'}^{\ast\{O\}}$ quantifies the contributions of operators of order $n'$ to $O$ via $\Shell_O$ and $b_{n=3+4}^{\{O\}} = b_{n=3}^{\{O\}} + b_{n=4}^{\{O\}}$.
We refer to Paper 1 for a broader discussion on the approximation that considers those higher-order operators as constants evaluated at $\L_\ast$, as well as the analytical solution for the bias RGE. In particular, it was pointed out that it is important to include higher-order operators, but sufficient to approximate their coefficients as constants. Therefore, in order to avoid cluttering the text, hereafter we omit terms such as the last terms in \refeqs{drun}{Grun}, that account for the truncation effect of approximating higher-order operators as constants.

\paragraph{$\bm{J^2}$: stochastic power spectra.}
The running of the stochastic parameters is considered here (to our knowledge) for the first time in the LSS literature. We consider the equation for the first and zeroth-order terms. Recall that $P_{\eps,\unit} = C_\unit^{(2)}$ enters in the power spectrum, cf. \refeq{dd_general}, while $P_{\eps,\d} = C_\d^{(2)}$ appears in the bispectrum, cf. \refeq{bispec}.
We find that 
\ba
\frac{d P_{\eps,\unit}}{d \L} &= - 2 [b_{\d^2}(\L)]^2 \left[{2}\Plin(\L)\right]  \frac{d\s^2_\L}{d\L} \,, \label{eq:Punitrun}
\\
\frac{d P_{\eps,\d}}{d \L} &= -\left[\frac{68}{21}P_{\eps,\d^2}^\ast+P_{\eps, n=3,\o=1}^{\ast\{\d\}} \right]\frac{d \s^2_\L}{d \L} \vs & \hspace{2cm}- \left[ \frac{136}{21} b_{\d^2}(\L)b_{\d^2}(\L) + P_{\eps, n=3,\o=2}^{\ast\{\d\}}  \right] \left[{2}\Plin(\L)\right]  \frac{d\s^2_\L}{d\L} \,, \label{eq:Pdrun}
\ea
where the terms $\propto\left[b_{\d^2}(\L)\right]^2$ are the novel sources from $\Shell^{22}_{\d^2\d^2}$.  Notice an important simplification taken here, which involves fixing $P_{\eps,\d^2}$ at $\L_*$. The complete evaluation of this term requires calculating $S^{22}_{\d^2\d^2}$ with two external legs, as well as contributions involving third and fourth-order operators, which is beyond the scope of this paper. We define $P^\ast_{\eps,O} = P_{\eps,O}(\L_\ast)$ and similar to \refeqs{hobd}{hobG2} we defined
\ba
P_{\eps, n=3,\o=1}^{\{\d\}} &=  3P_{\eps,\d^3}-\frac{4}{3}P_{\eps,\G_2\d} \,, \\
P_{\eps, n=3,\o=2}^{\{\d\}} &=  12 b_{\d^3}b_{\d^2} -\frac{16}{3} b_{\G_2\d}b_{\d^2} \,, 
\ea
to account for higher-order operators. Here, $P_{\eps, n=3,\o=1}^{\{\d\}}$ and $P_{\eps, n=3,\o=2}^{\{\d\}}$ account respectively for the $\Shell_{O_1}$, $\Shell_{O_1O_2}$ contributions to $\d$ that have at least one operator that is third-order or higher.  
We also included another subscript $\o$ to account for the number of operators in the contributing operator product. This is a generalization of the terms such as $b_{n=3}^{\{O\}}$ presented above, which includes contributions from multiple $\Shell_{O_1 \dots O_\o}$ which appear starting from order $J^p$.

\paragraph{$\bm{J^3}$: stochastic bispectra.}
The $J^3$ terms start to contribute at the bispectrum level.
We have
\ba
\frac{d B_{\eps,\unit}}{d \L} &=  - 2 P_{\eps,\d^2}^\ast b_{\d^2}(\L) \left[{2}\Plin(\L)\right]  \frac{d\s^2_\L}{d\L} -  8 [b_{\d^2}(\L)]^3 {3}\left[ \Plin(\L)\right]^2  \frac{d\s^2_\L}{d\L} \,, \label{eq:Bunitrun}
\\
\frac{d B_{\eps,\d}}{d \L} &= -\left[\frac{68}{21}B_{\eps,\d^2}^\ast+ B_{\eps, n=3,\o=1}^{\ast\{\d\}} \right]\frac{d \s^2_\L}{d \L}  - \left[ \frac{136}{21} P_{\eps,\d^2}^\ast b_{\d^2} + B_{\eps, n=3,\o=2}^{\ast\{\d\}} \right]\left[{2}\Plin(\L)\right]  \frac{d\s^2_\L}{d\L}    \label{eq:Bdrun}
\\ &\hspace{3cm}- \left[ \frac{544}{21} b_{\d^2}(\L)b_{\d^2}(\L)b_{\d^2}(\L) + B_{\eps, n=3,\o=3}^{\ast\{\d\}} \right] {3}\left[ \Plin(\L)\right]^2  \frac{d\s^2_\L}{d\L} \,, \nonumber
\ea
where we included the novel contribution from $\Shell^{222}_{\d^2\d^2\d^2}$ in the last lines of each ODE. We also define $B^\ast_{\eps,O} = B_{\eps,O}(\L_\ast)$ and
\ba
B_{\eps, n=3,\o=1}^{\{\d\}} &=  3B_{\eps,\d^3}-\frac{4}{3}B_{\eps,\G_2\d} \,, \\
B_{\eps, n=3,\o=2}^{\{\d\}} &=  6 P_{\eps,\d^3}b_{\d^2} -\frac{8}{3} P_{\eps,\G_2\d}  b_{\d^2} + 6 b_{\d^3}P_{\eps,\d^2} -\frac{8}{3} b_{\G_2\d} P_{\d^2} \,, \\
B_{\eps, n=3,\o=3}^{\{\d\}} &=  72 b_{\d^3}b_{\d^2}b_{\d^2}  - 32 b_{\G_2\d}b_{\d^2} b_{\d^2}  \,, 
\ea
to account for higher-order operators for the stochastic term. Here, $B_{\eps, n=3,\o=1}^{\{\d\}}$, $B_{\eps, n=3,\o=2}^{\{\d\}}$  and $B_{\eps, n=3,\o=3}^{\{\d\}}$  account respectively for the $\Shell_{O_1}$, $\Shell_{O_1O_2}$ and $\Shell_{O_1O_2O_3}$ contributions to $\d$ that have at least one operator that is third-order or higher.

\paragraph{Solutions and discussion.}
First, we note the vertical coupling of the RGE, where the equations for bias terms, i.e. order $J^1$, are independent of $J^2, J^3, ...$ coefficients (see \reffig{sourcestructure}), and the equations for the $J^2$ coefficients are independent of the $\O \left[J^3\right]$ terms.
This points to an iterative solution, starting with $J^1$, and a consistent way to truncate the RGE hierarchy in powers of $J$ (as first observed in  \cite{Carroll:2013oxa}). 
In addition to that truncation in $J$, which corresponds to a truncation of the set of $n$-point functions one is interested in, there are necessary truncations in terms of order of operators, and in terms of derivative orders $d_O$, which are suppressed by $\left(k/\L\right)^2$. These truncations, which are discussed in Section~2 of Paper 1, remain similarly valid for the stochastic terms as well. 

Note also that universal $m$-independent coefficients appear in the RGE; for example, the prefactor $68/21$ multiplying $C^{(m)}_{\d^2}$ in the RGE for $C^{(m)}_{\d}$, explicitly shown above for $m=1,2,3$ [see \refeq{drun}, \refeq{Pdrun} and  \refeq{Bdrun}], which is a direct consequence of the same interaction kernels appearing in the same kinematic configuration in these contributions. Unfortunately, this does not mean we can make general predictions for relations between the $C^{(m)}_\d$, for example, because different high-order operators contribute to the running for each $m$.

Moreover, whereas the bias parameters admit a variable change solving directly for $d b_O / d {\s^2_\L}$, the source terms for the stochastic parameters are controlled by $[\Plin(\L)]^{p} d\sigma_\L^2/d\L$ [see \refeq{masterSol}].
The velocity with which those parameters run in the RG-flow is thus determined by how large $\Plin(\L)$ is. 
An analytic solution to the RG-flow of terms starting from $J^2$ is not easily achievable due to the non-analytic form of $\Plin(\L)$. One could attempt to find solutions for power-law Universes or consider a \texttt{FFTLog} expansion of the $\Lambda$CDM matter power spectra for that \cite{McEwen:2016fjn,Simonovic:2017mhp}, which we leave for future work.

\begin{figure}[t]
	\centering
	\includegraphics[width = 0.45\textwidth]{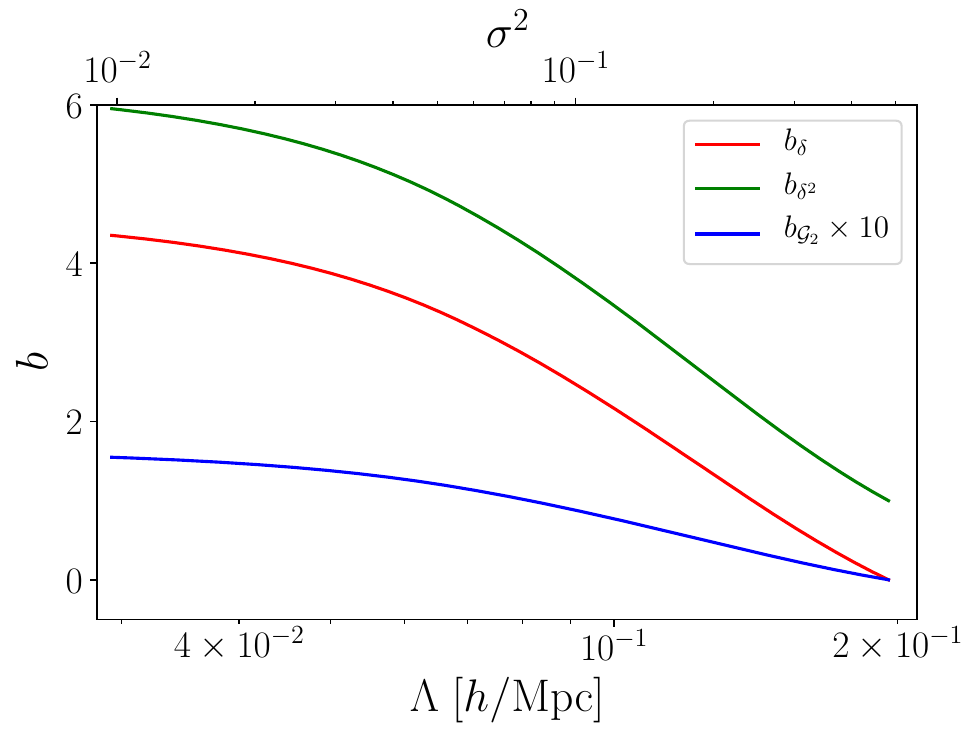}
	\includegraphics[width = 0.45\textwidth]{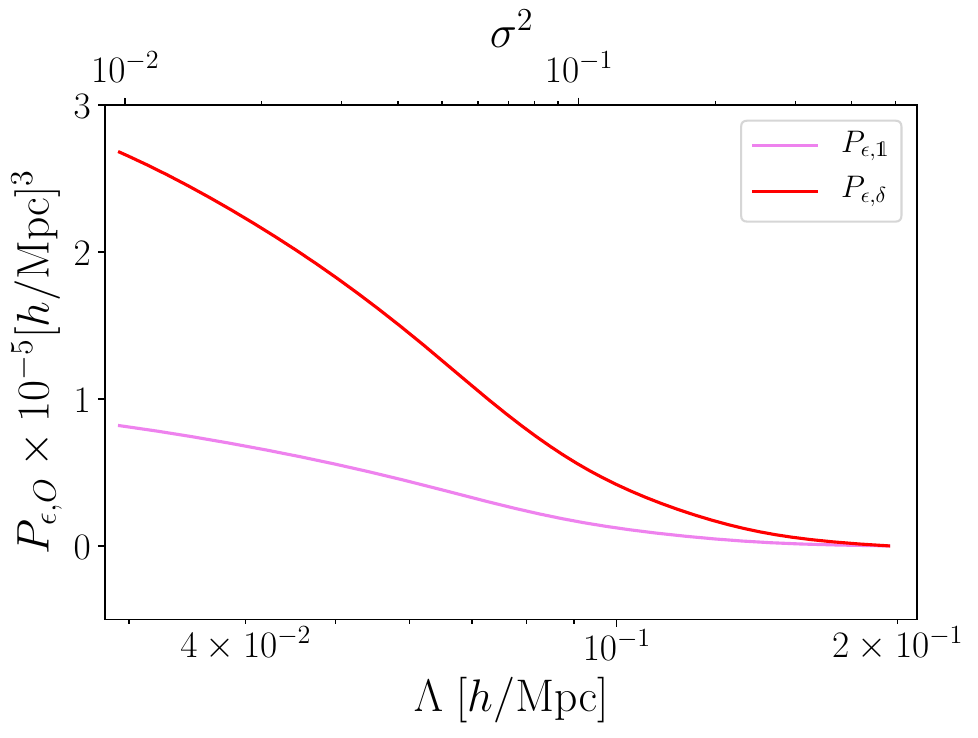}
	\includegraphics[width = 0.45\textwidth]{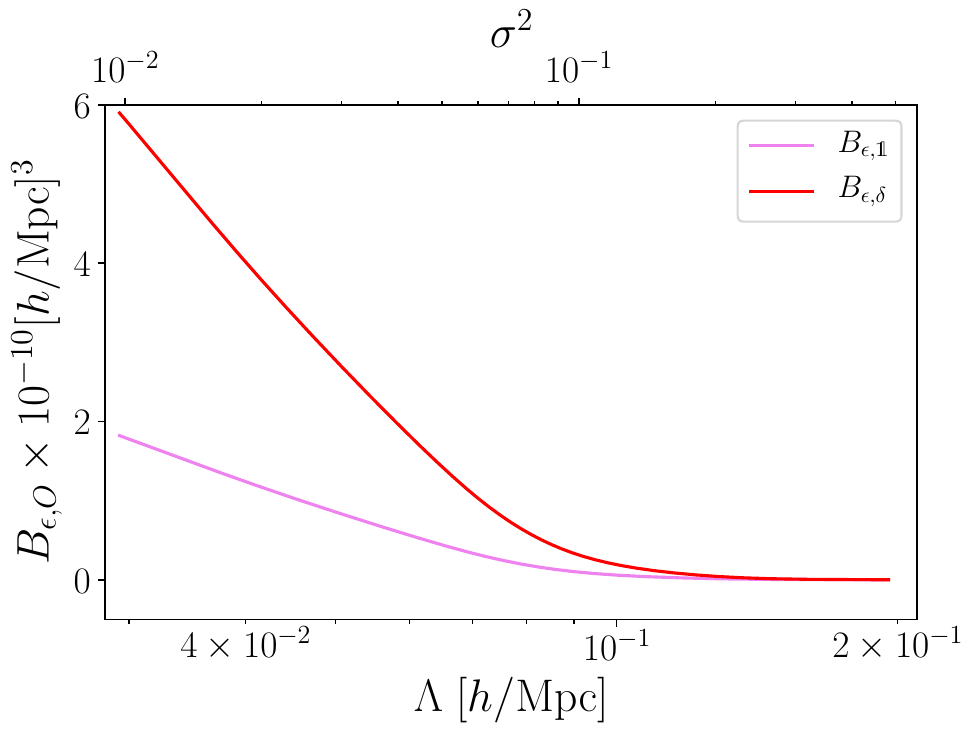}
	\caption{RG-flow for the bias (top left), stochastic (top right) and non-Gaussian stochastic (bottom) coefficients as a function of the cutoff $\L$. The initial conditions are set as $b_{\d^2}^\ast = 1$, $P_{\eps,\d^2}^\ast = \Plin(\L_\ast)$ and $B_{\eps,\d^2}^\ast = \left[\Plin(\L_\ast)\right]^2$ at $\s^2_\ast = 0.5$, with all other parameters set to zero on this scale. }
	\label{fig:ode_1}
\end{figure}

\begin{figure}[t]
    \centering
    \includegraphics[width = 0.7\textwidth]{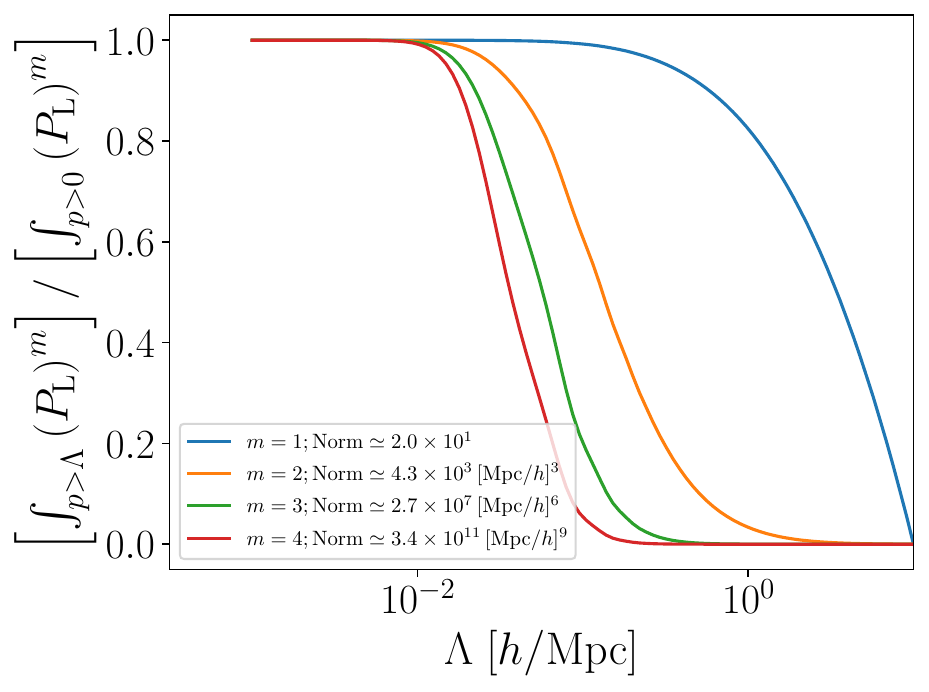}
    \caption{Expected RG evolution of the coefficients $C_O^{(m)}$ at different orders in $J^m$, as described by \refeq{lininlambda}. We normalize each line by the value when taking $\L \to 0$, $\int_{\vp>0}(\Plin)^m$, with their order of magnitude shown in the legend of the figure. Here, the upper limit of the integral is taken to be $10 \,h/{\rm Mpc}$; notice that the variance of the linear density ($m=1$) does depend on this choice.}
    \label{fig:Pscaling}
\end{figure}

In order to illustrate the RGE results, we consider the following scenario for the initial condition for the $\d^2$ parameters, that we have seen appear in the source of stochastic operators, in which we fix 
\ba
b_{\d^2}(\L_\ast) &= b_{\d^2}^\ast = 1\,, \\
P_{\eps,\d^2}(\L_\ast) &= P_{\eps,\d^2}^\ast = \Plin(\L_\ast)\,, \\
B_{\eps,\d^2}(\L_\ast) &= B_{\eps,\d^2}^\ast = \left[\Plin(\L_\ast)\right]^2\,,
\label{eq:ICexample}
\ea
at $\s^2_\ast = 0.5$ (corresponding to the renormalization scale $\L_\ast \approx 0.2 \,h/$Mpc), while all others are set to zero at the same scale:
\ba
b_{\d}(\L_\ast) = b_{\G_2}(\L_\ast) = P_{\eps,\d}(\L_\ast)= P_{\eps,\G_2}(\L_\ast) = B_{\eps,\d}(\L_\ast) = B_{\eps,\G_2}(\L_\ast) = 0\,.
\ea
That situation corresponds to a case in which ${\d^2}$ terms are responsible for populating all other operators. 
The solutions for this scenario are shown in \reffig{ode_1}. We show solutions for the $J^1$ (top left), $J^2$ (top right) and $J^3$ (bottom) parameters as a function of the scale $\L$. 
As discussed in Paper 1 and seen in \refeqs{drun}{Grun}, the running of $J^1$ terms, when neglecting higher-order operators, is sourced by $b_{\d^2}$. 
For the $J^2$ terms, notice that despite those parameters starting from zero at $\L = \L_\ast$, they rise sharply to $10^4-10^6 \,[h/{\rm Mpc}]^3$. These dynamics can be explained through the shape of the $\L$CDM matter power spectrum, together with \refeqs{Punitrun}{Pdrun}. Parametrically, when running from a high scale $\L_\ast$ to $\L \ll \L_\ast$, the change in $P_{\eps,O}$ is given by 
\be 
\Delta P_{\eps,O} \sim \int_{p>\L} \Plin^2(p) \stackrel{\L \to 0}{\approx} 4.3\times 10^3 \,\,[h/{\rm Mpc}]^3 \,,
\label{eq:PepsOrun}
\ee
for a $\L$CDM power spectrum, explaining the order of magnitude seen in the figure [note that the contribution is controlled by $b_{\d^2}(\L)$, which likewise grows toward smaller $\L$ and is not included in the estimate \refeq{PepsOrun}]. Similarly,
\be 
\Delta B_{\eps,O} \sim \int_{p>\L} \Plin^3(p) \stackrel{\L \to 0}{\approx} 2.7\times 10^{7} \,\, [h/{\rm Mpc}]^6\,.
\label{eq:BepsOrun}
\ee
The reason for the rapid running is the shape of the linear matter power spectrum: both of the above integrals are in fact dominated by the lowest wavenumbers, close to $\L$ (as long as $\L$ is greater than the turnover scale in the power spectrum, $\sim 0.02\iMpch$). A similarly enhanced contribution dominated by IR modes was pointed out for the matter four-point function in \cite{2017JCAP...11..051B}. This is in contrast to the corresponding integral appearing in the RGE for bias parameters, which is controlled by the variance of the linear density field, $\int_{p>\L} \Plin(p)$, and is dominated by UV modes. The dominance of low-$k$ modes in fact grows as one considers higher $m$, i.e. higher stochastic moments. This is illustrated in \reffig{Pscaling}. In hindsight, this behavior is not surprising: \refeqs{PepsOrun}{BepsOrun} give the expected order of magnitude of those operators
  since $\Plin(\L)$ is the only dimensionful quantity that appears in the RG equations at leading order in derivatives.

These results have interesting implications for the choice of the renormalization scale $\L_\ast$ when attempting to model the clustering of actual LSS tracers: choosing a low cutoff means that stochastic amplitudes will be enhanced, reducing the effective signal-to-noise of the measurement (recall that the stochastic terms also appear in the covariance, i.e. the likelihood of $n$-point functions). Thus, one should attempt to increase $\L_\ast$ to the highest scale still amenable to a perturbative description. We aim to investigate the optimal choice for the renormalization scale in a future work.

It is further worth highlighting that 
\be
\frac{dP_{\eps,\unit}}{d\L} \propto \left(b_{\d^2}\right)^2\,,
\ee
meaning that if at one fixed $\L_\ast$ we have $P_{\eps,\unit}\geq 0$,
as physically required since it corresponds to a variance,
  then this coefficient will remain positive for all $\L<\L_\ast$.
To see why this is physically required, consider the case of a tracer with vanishing $b_\d$. The power spectrum [\refeq{bispec}], which has to be non-negative, is then $\propto P_{\eps,\unit}$ on large scales.

\begin{figure}[t]
	\centering
	\includegraphics[width = 0.45\textwidth]{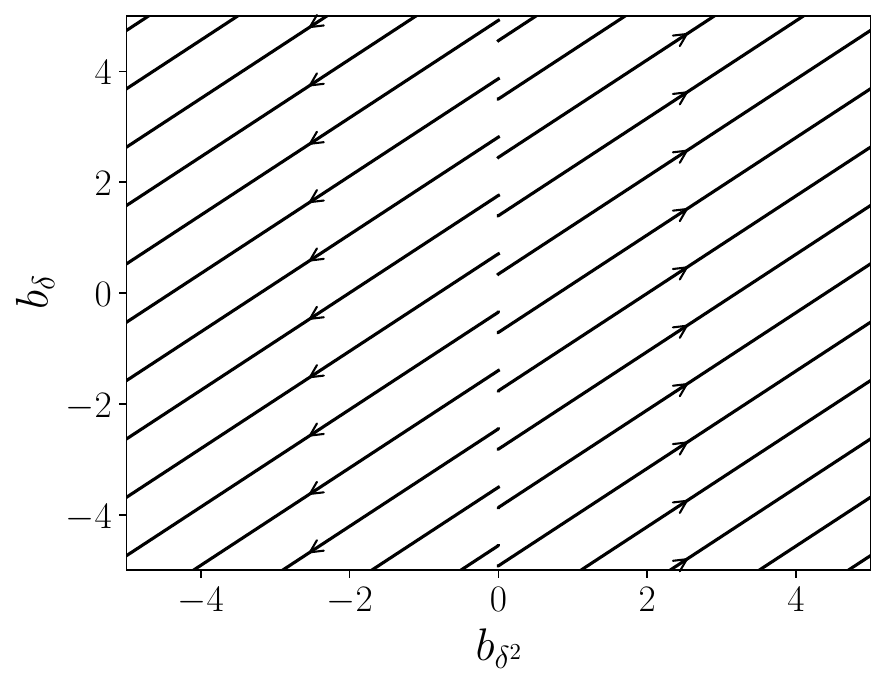}
	\includegraphics[width = 0.45\textwidth]{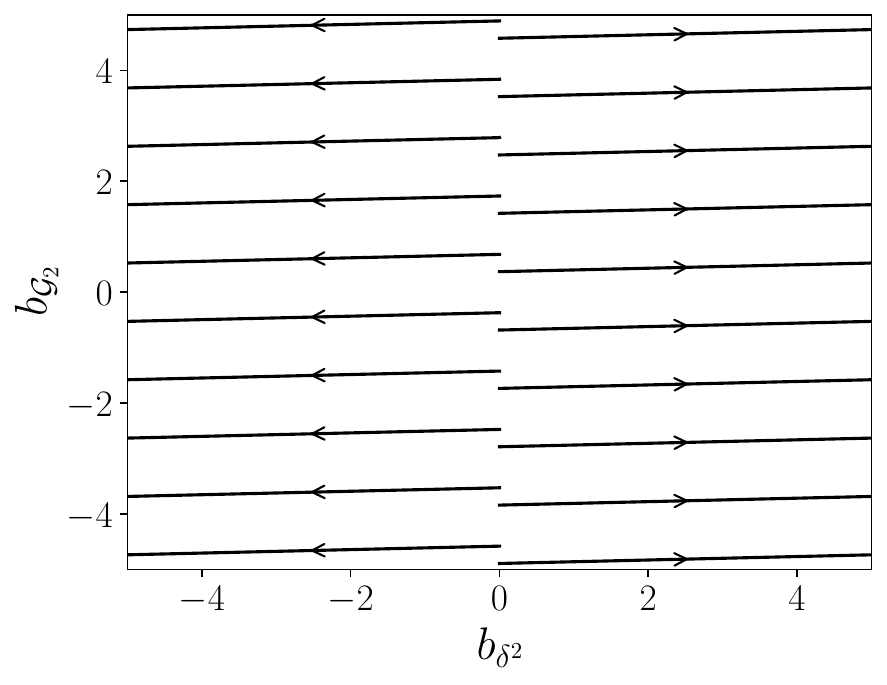}
	\includegraphics[width = 0.45\textwidth]{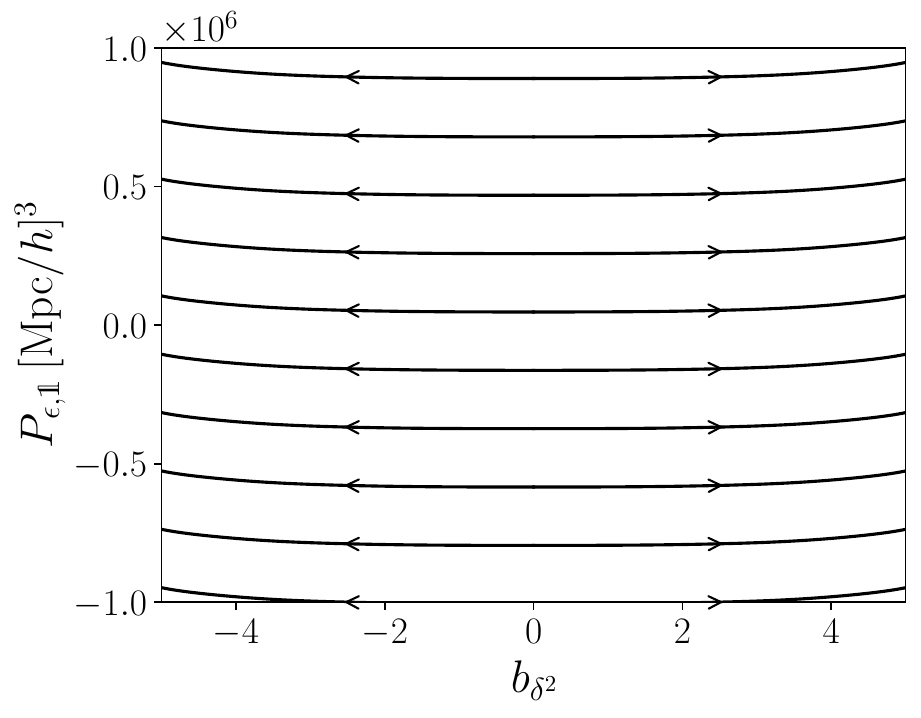}
	\includegraphics[width = 0.45\textwidth]{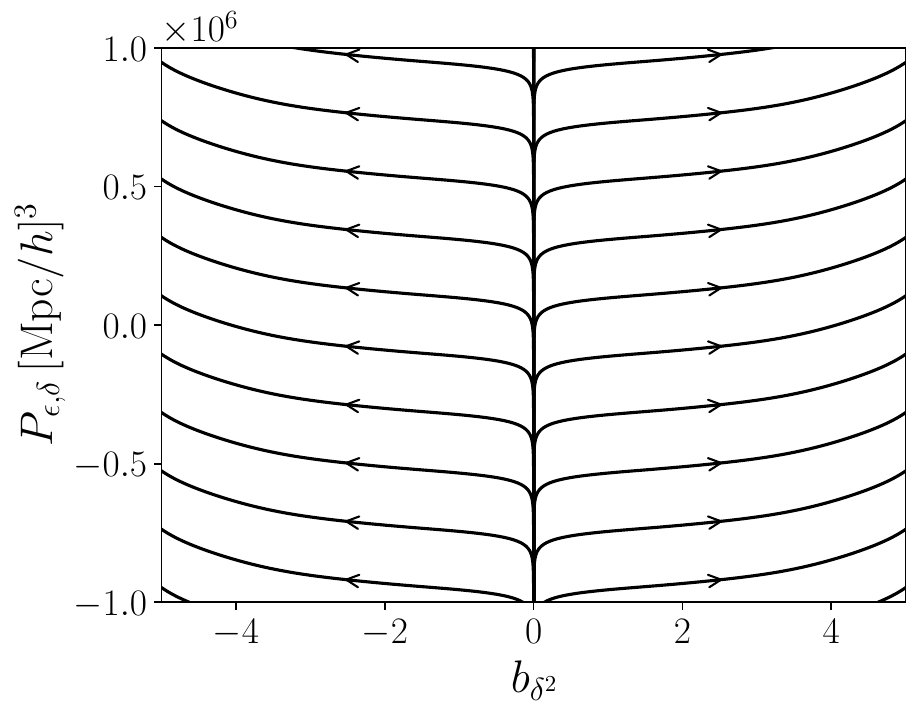}
	\includegraphics[width = 0.45\textwidth]{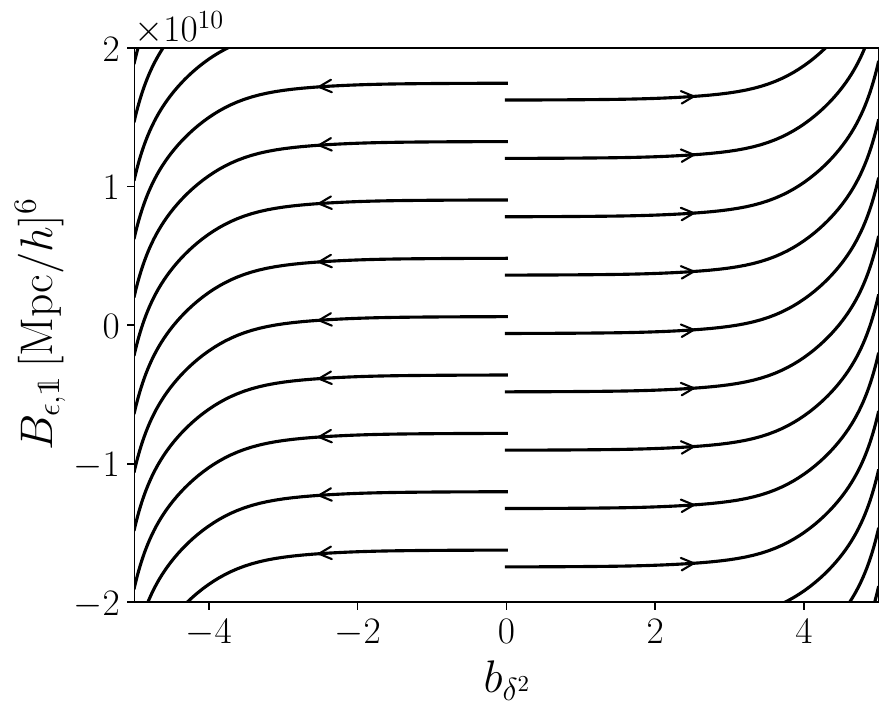}
	\includegraphics[width = 0.45\textwidth]{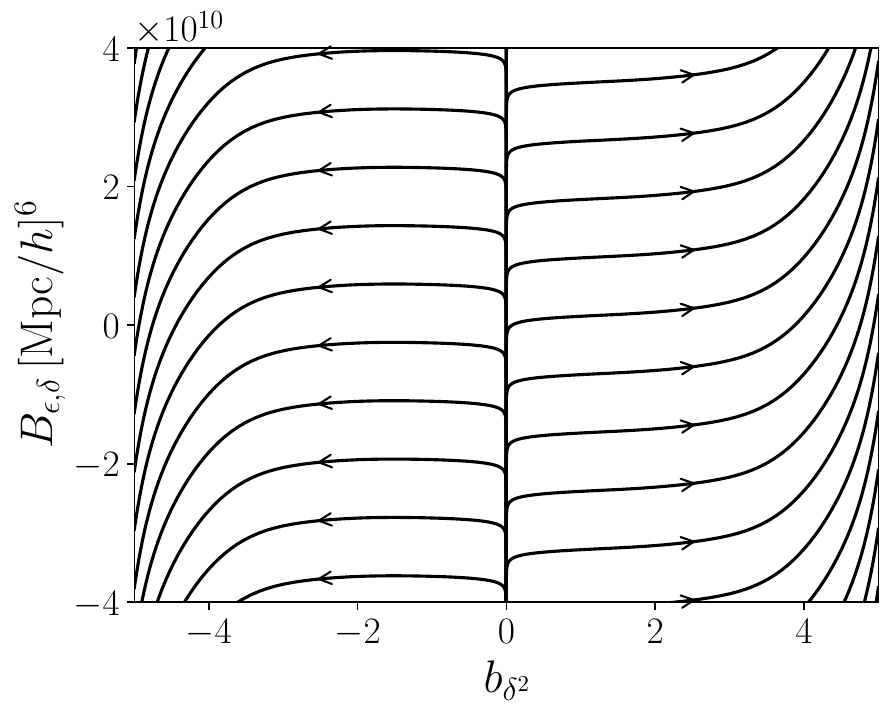}
	\caption{The RG-flow trajectories for different initial conditions in different parameter planes, always as a function of $b_{\d^2}$. The arrows indicate the running towards $\L \to 0$. Each line takes different initial conditions in the plane of parameters described by each figure, with $P_{\eps,\d^2}^\ast = \Plin(\L_\ast)$ and $B_{\eps,\d^2}^\ast = \left[\Plin(\L_\ast)\right]^2$ and the other parameters (but for those shown in each panel) fixed to zero in the limit $\L_\ast \to 0$. }
	\label{fig:rgflow_1}
\end{figure}

Let us now move beyond the simple initial conditions considered for \reffig{ode_1}, \refeq{ICexample}.
We display in \reffig{rgflow_1} the RG-flow of different parameters as a function of $b_{\d^2}$. 
The top, middle and bottom panels present respectively the $J^1$, $J^2$ and $J^3$ terms. Here, we focus on a 2-d slice in the multidimensional parameter space that describes the running. 
Each line points towards the direction $\L \to 0$, with different initial conditions in the respective $\{b_{\d},\,b_{\G_2},\,P_{\eps,\unit},\,P_{\eps,\d},\,B_{\eps,\unit},\,B_{\eps,\d} \} \times b_{\d^2}$ planes shown, with $P_{\eps,\d^2}^\ast = \Plin(\L_\ast)$ and $B_{\eps,\d^2}^\ast = \left[\Plin(\L_\ast)\right]^2$  and the other parameters, not shown in the respective figures, set to zero in the limit $\L_\ast \to 0$. 
Note that the running of $b_\d, b_{\d^2}$ is much stronger than that of $b_{\G_2}$, as expected and already shown in Fig.~2 of Paper 1.  
For the running of the $P_{\eps,O}$ in the second row, presented in this work for the first time, we note especially the inflection point visible for $P_{\eps,\d}$, which can be understood as a competition between the $\left(b_{\d^2}\right)^2$ (from $\Shell^{22}_{\d^2\d^2}$) and $P_{\eps,\d^2}$ (from $\Shell^{2}_{\d^2}$) in \refeq{Pdrun}. For the $J^3$ terms in the last row, we see a complex structure in the RG-flow emerging from the different terms in \refeq{Bunitrun} and \refeq{Bdrun}.


\section{Conclusion}
\label{sec:conc}

  We have used the general effective action for large-scale structure, \refeq{Sreal}, to derive the ``RG-LSS'' equations which describe the running of bias and stochastic contributions to the clustering of LSS tracers under a change of the renormalization scale. This extends the analysis of Paper 1 \cite{Rubira:2023vzw}, which focused on bias, i.e. terms linear in the current $J$, to arbitrary powers of $J$. 
  Our result therefore is the most general RG running for the leading-in-derivative terms of the EFT of LSS.
In \refsec{running}, we make use of the Wilson-Polchinski formalism to derive how those terms are sourced by bias operators and lower-order-in-$J$ stochastic parameters. We provide a general master formula \refeq{masterSol} for the evolution of any stochastic or bias parameter as a function of the cutoff $\L$, which simplifies to \refeq{dUnit} for the shot-noise contributions to $n$-point functions. 
We also analyse in \refsec{results} solutions for the RG-flow including stochastic parameters.

In this work we continue to follow the philosophy introduced in Paper 1, which keeps the smoothing cutoff (i.e., the renormalization scale) finite, instead of subtracting the leading order contributions by taking the large-scale limit $\L \to 0$ as considered in usual EFT of LSS analyses (see Sections 3 and 4 of Paper 1 for a comparison between both finite-$\L$ and $n$-point-function renormalization schemes).\footnote{For the renormalization of stochastic terms, see e.g. Appendix D of \cite{DAmico:2022osl,DAmico:2022ukl}.} 
The finite-$\L$ approach allows for the derivation of the RG equations using the Wilson-Polchinski framework, which may account for a yet-to-be-determined amount of extra information from the resummation of part of higher-loop contributions as in the usual context of quantum field theory, where large logarithms are resummed. 
We find that, different than for the bias parameters, the RG equations for the stochastic parameters are nonlinear, which may suggest that some relevant information may be absorbed by their RG flow. The discussion of whether more (non-perturbative) information can be extracted from the RG equations for the LSS coefficients is deferred for a future publication.  
Furthermore, non-trivial critical points may arise from the RG flow structure of the bias and stochastic parameters.
These are directions we aim to investigate in the future.
Another interesting generalization to be understood in light of the RG equations is the presence of primordial non-Gaussianities \cite{Nikolis:2024kbx}. 
Non-Gaussianities lead to new types of vertices for which the RG equations (and their inbuilt resummation) may be relevant. 


\acknowledgments
HR is supported by the Deutsche Forschungsgemeinschaft under Germany's Excellence Strategy EXC 2094 `ORIGINS'. (No.\,390783311). 
We thank Giovanni Cabass, Mathias Garny, Matt Lewandowski, Mehrdad Mirbabayi, Charalampos Nikolis, and Marko Simonovic for discussions and Charalampos Nikolis and Beatriz Tucci for feedback on the paper. 

\appendix

\section{Evaluation of shell integrals including relevant contributions for the stochastic terms}  \label{app:shell}

In this appendix, we focus on calculating the shell integrals $\Shell_{O}$, $\Shell_{OO'}$ and $\Shell_{OO'O''}$. 
We direct the reader to Appendix~A.1 of Paper 1 for a complete form of the shell operators \refeq{Oshell}. 

\subsection{The  \texorpdfstring{$\Shell_{O}^{2}$}{SO2} integrals} \label{app:S2eval}
We start reviewing part of the results of Appendix~A of Paper 1, in which 
\ba
\Shell_{O}^{2} [\dlin_\L](\vk) &= \sum_{n\geq 2} \int \Del\dlinshell \P[\dlinshell] \: O^{(n),(2)_{\rm shell}} [\dlin_\L,\dlinshell](\vk) 
\vs
&= \sum_{n\geq 2} \left\< O^{(n),(2)_{\rm shell}} [\dlin_\L,\dlinshell](\vk) \right\>_{\rm shell} \,,
\label{eq:shellrepeat}
\ea
is calculated. We split the calculation in terms of the $\ell$ number of external legs 
\ba
\Shell_{O}^{2} [\dlin_\L] (\vk) &=  \sum_{\ell \geq 0} \left(\Shell_{O}^{2} [\dlin_\L] \right)_{(\ell)_{\rm legs}} (\vk)\,,
\label{eq:nlegs}
\ea
in which, e.g., $\ell = 0$ corresponds to corrections to $C^{(m)}_{\unit}$, $\ell = 1$ corresponds to corrections to $C^{(m)}_{\d}$, $\ell = 2$ corresponds to corrections to second-order operators as 
\ba
    \left(\Shell_{O}^{2} [\dlin_\L] \right)_{(\ell)_{\rm legs}}(\vk) = \raisebox{-0.0cm}{\includegraphicsbox[scale=.9]{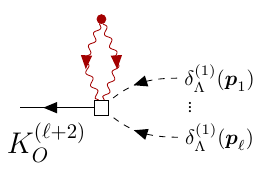}} \,. \nonumber
\ea

\paragraph{Contributions from $\left(\Shell_{O}^2\right)_{(0)_{\rm legs}}$.}
The $\d^2$ contribution leads to 
\be
\left(\Shell_{\d^2}^2\right)_{(0)_{\rm legs}}(\vk) =  \diracpi(\vk)  \int_{\vp} P_{\rm shell}(p)\,,
\ee
and then, e.g.,
\bea
&&\frac12 P_{\eps,\d^2}({\L}) \int_{\vk_1,\vk_2,\vk_3} \diracpi(\vk_{123}) J_{\L}(\vk_1) J_{\L}(\vk_2)    \left(\Shell_{\d^2}^2\right)_{(0)_{\rm legs}} (\vk_3)  \vs
&& \hspace{2cm} = \frac12 P_{\eps,\d^2}({\L})\left[\int_{\vp} P_{\rm shell}(p)\right] \int_{\vk_1,\vk_2} \diracpi(\vk_{12}) J_{\L}(\vk_1) J_{\L}(\vk_2) \,, 
\eea
which would contribute to $P_{\eps,\unit}$. This contribution, however, is removed after the normalization \refeq{Onorm}. The other two zero-leg contributions come from $\d$ and $\G_2$ which lead to
\bea
\left(\Shell_{\d}^2\right)_{(0)_{\rm legs}}(\vk) &=&  \diracpi(\vk)  \int_{\vp} F_2(\vp,-\vp)P_{\rm shell}(p) = 0\,, \label{eq:Shell_d0}\\
\left(\Shell_{\G_2}^2\right)_{(0)_{\rm legs}}(\vk) &=&  \diracpi(\vk)  \int_{\vp} \s^2_{\vp,-\vp} P_{\rm shell}(p) = 0 \label{eq:Shell_G20}\,.
\eea

\paragraph{Contributions from $\left(\Shell_{O}^2\right)_{(1)_{\rm leg}}$.}
The only terms that contribute to the running of leading in derivatives single-leg operators are those from $\d^2$, $\d\G_2$ and $\d^3$ shell-integrals
\bea
\left(\Shell_{\d^2}^2\right)_{(1)_{\rm leg}}(\vk) &=& 4 \dlin_\L(\vk) \int_{\vp }F_2(\vk, \vp)P_{\rm shell}(p) 
=  \frac{68}{21} \dlin_\L(\vk) \int \frac{p^2dp}{2\pi^2}P_{\rm shell}(p)\,, \\
\left(\Shell_{\d^3}^2\right)_{(1)_{\rm leg}}(\vk) &=& 3 \dlin_\L(\vk) \int_{\vp }P_{\rm shell}(p)\,, \\
\left(\Shell_{\G_2\d}^2\right)_{(1)_{\rm leg}}(\vk) &=& 2 \dlin_\L(\vk) \int_{\vp }\s^2_{\vk,\vp}P_{\rm shell}(p)  = - \frac{4}{3} \dlin_\L(\vk)  \int \frac{p^2dp}{2\pi^2}P_{\rm shell}(p)\,.
\eea

\paragraph{Contributions from $\left(\Shell_{O}^2\right)_{(2)_{\rm legs}}$.}

Omitting terms that only contribute to higher-derivative operators, we find  
\bea \label{eq:Sd2A}
\left(\Shell^2_{\d^2}\right)_{(2)_{\rm legs}}(\vk)  &=& \left[ \frac{68}{21}\d^{(2)}(\vk) + \frac{8126}{2205}\left[\d^{2}(\vk)\right]^{(2)} +\frac{254}{2205}\G_2^{(2)}(\vk) \right] \int \frac{p^2 dp}{2 \pi^2 }P_{\rm shell}(p)\,, \label{eq:S2d2calc}\\
\left(\Shell^2_{\d^3} \right)_{(2)_{\rm legs}}(\vk) &=&  3 \left(\d[\d_\L^{(1)}]\right)^{(2)}(\vk)  \int_{\vp} P_{\rm shell}(p) + \frac{68}{7}  \left(\d^2[\d_\L^{(1)}]\right)^{(2)}(\vk)   \int_{\vp}   P_{\rm shell}(p)  \,, \\
\left( \Shell^2_{\G_2\d}\right)_{(2)_{\rm legs}}(\vk) &=& \left[ - \frac{4}{3}\d^{(2)}(\vk) - \frac{376}{105}\d^{2,(2)}(\vk) +\frac{116}{105}\G_2^{(2)}(\vk) \right] \int \frac{p^2 dp}{2 \pi^2 }P_{\rm shell}(p) \,.
\eea
Notice that part of those terms also contribute to one-leg terms found above with the same coefficients, as a consequence of the equivalence principle \cite{Rubira:2023vzw}.

Fourth-order operators also introduce contributions of the type $\left(\Shell_{O}^2\right)_{(2)_{\rm legs}}$, e.g.
\bea
\left(\Shell^2_{\d^4} \right)_{(2)_{\rm legs}}(\vk) &=&  6\left(\d^2[\d_\L^{(1)}]\right)^{(2)}(\vk)\int_{\vp} P_{\rm shell}(p)  \,.
\eea
We leave the calculation of the full contribution of the set of fourth-order terms to a future project.

\subsection{The suppression of \texorpdfstring{$\left(\Shell_{O}^m\right)_{(\ell)_{\rm legs}}$}{SOmllegs} for \texorpdfstring{$m>2$}{m>2} } \label{app:S4suppression}

In this part we discuss how higher-loop contributions such as  $\left(\Shell_{O}^4\right)_{(\ell)_{\rm legs}}$ [see \reffig{SOdiag} for a diagrammatic representation] and $\left(\Shell_{O}^6\right)_{(\ell)_{\rm legs}}$ are suppressed compared to  $\left(\Shell_{O}^2\right)_{(\ell)_{\rm legs}}$ . 
In order to exemplify that, we consider the contributions $\left(\Shell_{O}^4\right)_{(0)_{\rm legs}}$, which correct the zeroth-leg operator. Notice that also third and fourth-order operator $O$ also contribute to $\left(\Shell_{O}^4\right)_{(0)_{\rm legs}}$, but analyzing the first and second-order operators contributions is enough to understand the general structure of the $\left(\Shell_{O}^4\right)_{(0)_{\rm legs}}$:
\ba
\left(\Shell_{\d}^4\right)_{(0)_{\rm legs}}(\vk) &=  3 \diracpi(\vk)  \int_{\vp_1, \vp_2} F_4(\vp_1,-\vp_1,\vp_2,-\vp_2)P_{\rm shell}(p_1)P_{\rm shell}(p_2) \,, 
\\
\left(\Shell_{\d^2}^4\right)_{(0)_{\rm legs}}(\vk) & =  2 \diracpi(\vk)  \int_{\vp_1, \vp_2}\left[F_2(\vp_1, \vp_2)\right]^2  P_{\rm shell}(p_1) P_{\rm shell}(p_2) 
\\ 
& + 6 \diracpi(\vk)  \int_{\vp_1, \vp_2} F_3(\vp_1,-\vp_1, \vp_2)   P_{\rm shell}(p_1) P_{\rm shell}(p_2)\,,
\vs
\left(\Shell_{\G_2}^4\right)_{(0)_{\rm legs}}(\vk) &=  0 \,.
\ea
We can then summarize this type of contribution as 
\be
\left(\Shell_{O}^4\right)_{(0)_{\rm legs}}(\vk) \propto \diracpi(\vk)  \int_{\vp,\vp'} K_O^{(4)}(\vp, -\vp, \vp',-\vp') P_{\rm shell}(p) P_{\rm shell}(p')  \propto \sigma_{\rm shell}^4 \propto \lambda^2 ,
\ee
for a generic kernel $K_O^{(4)}$, where $\sigma_{\rm shell}^2 = d\sigma^2_\L/d\L\cdot\leps$. We can then easily generalize to write
\be
\left(\Shell_{O}^m\right)_{(0)_{\rm legs}}(\vk) \propto \sigma_{\rm shell}^m \propto \lambda^{m/2} ,
\ee
which has its leading contribution for $m=2$ and the others are suppressed by the shell width $\lambda$.
Similarly, we can summarize the contributions of $\left(\Shell_{O}^4\right)_{(\ell)_{\rm legs}}$ to $\ell$-legs operators as 
\ba
\int_{\vp,\vp'} K_O^{(\ell+4)}(\vk_1, \dots, \vk_\ell, \vp, -\vp, \vp',-\vp') P_{\rm shell}(p) P_{\rm shell}(p') \dlin_\L(\vk_1)\dots \dlin_\L(\vk_\ell) 
\vs
\hspace{2cm}\propto \sigma_{\rm shell}^4 \dlin_\L(\vk_1)\dots \dlin_\L(\vk_\ell) \propto \lambda^2 \dlin_\L(\vk_1)\dots \dlin_\L(\vk_\ell),
\ea
which is again suppressed.

\subsection{The  \texorpdfstring{$\Shell_{OO'}$}{SO} integrals} \label{app:SOOcalc}

We now move to calculate the $\Shell_{OO'}$ integrals, defined as
\ba
\Shell_{OO'}^{ii'} [\dlin_\L] (\vk,\vk') &= \sum_{n\geq i,n'\geq i'} \int \Del\dlinshell \P[\dlinshell] \: O^{(n),(i)_{\rm shell}} [\dlin_\L,\dlinshell](\vk)\: O'^{(n'),(i')_{\rm shell}} [\dlin_\L,\dlinshell](\vk') 
\vs
&= \sum_{n\geq i,n'\geq i'} \left\< O^{(n),(i)_{\rm shell}} [\dlin_\L,\dlinshell](\vk)\: O'^{(n'),(i')_{\rm shell}} [\dlin_\L,\dlinshell](\vk') \right\>_{\rm shell} \,.
\ea
We again split according to the $\ell$ number of external legs 
\ba
\Shell_{OO'}^{ii'} [\dlin_\L] (\vk,\vk') &=  \sum_{\ell\geq 0} \left( \Shell_{OO'}^{ii'} [\dlin_\L]  \right)_{(\ell)_{\rm legs}}(\vk,\vk')\,,
\nonumber
\ea
with 
\ba
    \left( \Shell_{OO'}^{22} [\dlin_\L]  \right)_{(\ell)_{\rm legs}}(\vk,\vk') = \raisebox{-0.0cm}{\includegraphicsbox[scale=0.9]{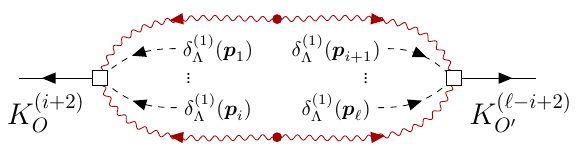}} \,. \vs
\ea
As we discuss in \refapp{J2structure}, the $\Shell^{11}$ and $\Shell^{13}$ terms are kinematically suppressed after considering the full partition function integrated with the currents $J_\L$. Despite they not contributing we still show part of those terms both for completeness and because we use them in \refapp{doubleshell}. We also only focus on the connected diagrams, since disconnected shell graphs do not contribute to the RG running.

\paragraph{Contributions from $\left(\Shell_{OO'}^{11}\right)_{(0)_{\rm legs}}$.}
The only zero-leg contribution of the type $\left(\Shell_{OO'}^{11}\right)_{(0)_{\rm legs}}$ comes from 
\bea
\left(\Shell^{11}_{\d\d}\right)_{(0)_{\rm legs}} (\vk,\vk') &=&  \diracpi(\vk+\vk')P_{\rm shell}(k) \,,
\eea
and is orthogonal to the external current $J_\L$
\bea
&& \frac12 b_\d({\L}) b_{\d}({\L}) \int_{\vk,\vk'}  J_{\L}(\vk) J_{\L}(\vk')   \left(\Shell^{11}_{\d\d}\right)_{(0)_{\rm legs}} (\vk,\vk') \vs
&& \hspace{2cm} = \frac12 b_\d({\L}) b_{\d}({\L}) \int_{\vk,\vk'} J_{\L}(\vk) J_{\L}(\vk') \diracpi(\vk+\vk')P_{\rm shell}(k) = 0\,. \label{eq:orthoconditionzeroleg}
\eea

\paragraph{Contributions from $\left(\Shell_{OO'}^{13}\right)_{(0)_{\rm legs}}$.}
All the zero-leg contributions of the type $\left(\Shell_{OO'}^{13}\right)_{(0)_{\rm legs}}$ will satisfy
\be
\left(\Shell_{OO'}^{13}\right)_{(0)_{\rm legs}}(\vk,\vk') \propto P_{\rm shell}(k) \,,
\ee 
e.g.:  
\bea
\left(\Shell^{13}_{\d\d}\right)_{(0)_{\rm legs}} (\vk,\vk') &=&  3 \diracpi(\vk+\vk')P_{\rm shell}(k)\int_{\vp} F_3(\vk, \vp, -\vp)P_{\rm shell}(p) \,.
\eea
As such, it will again be zero due to the orthogonality w.r.t the current $J_\L$.

\paragraph{Contributions from $\left(\Shell_{OO'}^{22}\right)_{(0)_{\rm legs}}$.}
The contributions of the type $\left(\Shell_{OO'}^{22}\right)_{(0)_{\rm legs}}$ are the following:
\ba
\left(\Shell^{22}_{\d\d}\right)_{(0)_{\rm legs}} (\vk,\vk') &=   2\diracpi(\vk+\vk')\int_{\vp} \left[ F_2(\vp, \vk-\vp) \right]^2 P_{\rm shell}(p) P_{\rm shell}(|\vk-\vp|) \,, \\
\left(\Shell^{22}_{\d\d^2}\right)_{(0)_{\rm legs}} (\vk,\vk') &=   2\diracpi(\vk+\vk')\int_{\vp} F_2(\vp, \vk-\vp) P_{\rm shell}(p) P_{\rm shell}(|\vk-\vp|) \,, \\
\left(\Shell^{22}_{\d^2\d^2}\right)_{(0)_{\rm legs}} (\vk,\vk') &=   2 \diracpi(\vk+\vk')\int_{\vp}  P_{\rm shell}(p) P_{\rm shell}(|\vk-\vp|) \label{eq:S22_d2d2_calc} \,.
\ea
Notice that the first two terms only contribute to higher-derivative operators, since $ F_2(\vp, \vk-\vp)  \propto k^2/p^2$ when $p \gg k$. The contributions involving $\G_2$ will also lead to higher-derivative contributions.

\paragraph{Contributions from $\left(\Shell_{OO'}^{11}\right)_{(1)_{\rm leg}}$.}
The one-leg contribution of the type $\left(\Shell_{OO'}^{11}\right)_{(1)_{\rm legs}}$ are 
\bea
\left(\Shell^{11}_{\d\d}\right)_{(1)_{\rm leg}} (\vk,\vk') &=&  4 P_{\rm shell}(k)  F_2(\vk + \vk', -\vk)\dlin_\L(\vk + \vk') \,, \\
\left(\Shell^{11}_{\d\d^2}\right)_{(1)_{\rm leg}} (\vk,\vk') &=&   2 P_{\rm shell}(k) \dlin_\L(\vk + \vk') +  \diracpi(\vk')\dlin_\L(\vk)  \int_{\vp} P_{\rm shell}(p) \,,\\
\left(\Shell^{11}_{\d\G_2}\right)_{(1)_{\rm leg}} (\vk,\vk') &=&   2 P_{\rm shell}(k) \s^2_{\vk+\vk',-\vk}\dlin_\L(\vk + \vk') \,.
\eea
Notice that all terms proportional to $P_{\rm shell}(k)$ will be zero when integrated with $J_\L(\vk)$ due to the orthogonality condition \refeq{orthocondition}. 

\paragraph{Contributions from $\left(\Shell_{OO'}^{22}\right)_{(1)_{\rm leg}}$.}
The only non-suppressed one-leg contributions of the type $\left(\Shell_{OO'}^{22}\right)_{(1)_{\rm legs}}$ are
\bea
\left(\Shell^{22}_{\d^2\d^2}\right)_{(1)_{\rm leg}} (\vk,\vk') &=& 8 \dlin_\L(\vk+\vk') \int_{\vp}F_2(\vk+\vk', -\vp)P_{\rm shell}(p)P_{\rm shell}(|\vk - \vp|)   \,,
\label{eq:S22d2d2oneleg} \\
\left(\Shell^{22}_{\d^3\d^2}\right)_{(1)_{\rm leg}} (\vk,\vk') &=& 6 \dlin_\L(\vk+\vk') \int_{\vp}P_{\rm shell}(p)P_{\rm shell}(|\vk - \vp|)   \,, \label{eq:S22d3d2oneleg}\\
\left(\Shell^{22}_{(\G_2\d)\d^2}\right)_{(1)_{\rm leg}} (\vk,\vk') &=& 4 \dlin_\L(\vk+\vk') \int_{\vp}\s^2_{-\vp,\vk+\vk'}P_{\rm shell}(p)P_{\rm shell}(|\vk - \vp|)   \,,
\label{eq:S22G2d2oneleg} 
\eea 
where we omitted the contribution that leads to higher-derivative operators. Notice here that third-order operators can also contribute. 

\paragraph{Contributions from $\left(\Shell_{OO'}^{11}\right)_{(2)_{\rm leg}}$.}

We find
\bea
\left(\Shell^{11}_{\d\d}\right)_{(2)_{\rm leg}} (\vk,\vk') =4 \int_{\vp}P_{\rm shell}(p) F_2(\vp,\vk-\vp)  F_2(-\vp,\vk'+\vp) \dlin_\L(\vk-\vp) \dlin_\L(\vk'+\vp)  \,. \,\,\,
\eea
However, the $F_2(\vp,\vk-\vp)$ kernel at $k\ll p$ scales as $k^2/p^2$, indicating that this term will source higher-derivative stochastic contributions. For the contraction of $\d$ with second-order operators we find
\bea
\left(\Shell^{11}_{\d\d^2}\right)_{(2)_{\rm leg}} (\vk,\vk') &=& 4 \int_{\vp}P_{\rm shell}(p) F_2(\vp,\vk-\vp)  \dlin_\L(\vk-\vp) \dlin_\L(\vk'+\vp) 
\vs 
&+&4 P_{\rm shell}(k) \int_{\vp_1,\vp_2} \diracpi(\vk + \vk'-\vp_{12}) F_2(-\vk,\vp_2)  \dlin_\L(\vp_1) \dlin_\L(\vp_2) 
\vs 
&+&2 P_{\rm shell}(k) \int_{\vp_1,\vp_2} \diracpi(\vk + \vk'-\vp_{12}) F_2(\vp_1,\vp_2)  \dlin_\L(\vp_1) \dlin_\L(\vp_2) \,, 
\vs
\left(\Shell^{11}_{\d\G_2}\right)_{(2)_{\rm leg}} (\vk,\vk') &=& 4 \int_{\vp}P_{\rm shell}(p) F_2(\vp,\vk-\vp) \s^2_{\vp,-\vk'-\vp}  \dlin_\L(\vk-\vp) \dlin_\L(\vk'+\vp)  
\vs 
&+&4 P_{\rm shell}(k) \int_{\vp_1,\vp_2} \diracpi(\vk + \vk'-\vp_{12}) F_2(-\vk,\vp_2) \s^2_{\vp_1,\vp_2-\vk} \dlin_\L(\vp_1) \dlin_\L(\vp_2) 
\vs 
&+&2 P_{\rm shell}(k) \int_{\vp_1,\vp_2} \diracpi(\vk + \vk'-\vp_{12}) F_2(\vp_1,\vp_2) \s^2_{\vp_1+\vp_2,-\vk}  \dlin_\L(\vp_1) \dlin_\L(\vp_2) \,, \nonumber
\eea
and the contraction of second-order with another second-order operators leads to
\bea
\left(\Shell^{11}_{\d^2\d^2}\right)_{(2)_{\rm leg}} (\vk,\vk') &=& 4 \int_{\vp}P_{\rm shell}(p)  \dlin_\L(\vk-\vp) \dlin_\L(\vk'+\vp)\,,
\label{eq:S11d2d2} \\
\left(\Shell^{11}_{\G_2\G_2}\right)_{(2)_{\rm leg}} (\vk,\vk') &=& 4 \int_{\vp}P_{\rm shell}(p) \s^2_{\vp,-\vk-\vp}\s^2_{\vp,-\vk'-\vp}  \dlin_\L(\vk-\vp) \dlin_\L(\vk'+\vp) \,.
\eea 
Notice that $\s^2_{\vp,-\vk-\vp}$ will also source higher-derivative terms. Terms proportional to $P_{\rm shell}(k)$ will again be zero when integrated with $J_\L(\vk)$ due to the orthogonality condition \refeq{orthocondition}. The only non-zero contribution at this level is then $\left(\Shell^{11}_{\d^2\d^2}\right)_{(2)_{\rm leg}}$, but it does not contribute in the limit of soft external momenta by the arguments given in \refapp{doubleshell}.

\subsection{The  \texorpdfstring{$\Shell_{OO'O''}$}{SOOO} integrals} \label{app:SOOOeval}

\paragraph{Contributions from $\left(\Shell_{OO'O''}^{222}\right)_{(0)_{\rm legs}}$.}
The contributions of the type $\left(\Shell_{OO'O''}^{112}\right)_{(0)_{\rm legs}}$,\\ $\left(\Shell_{OO'O''}^{114}\right)_{(0)_{\rm legs}}$ and $\left(\Shell_{OO'O''}^{123}\right)_{(0)_{\rm legs}}$ will have a term proportional to $P_{\rm shell}(k)$ and therefore will be zero due to the orthogonality condition w.r.t the current $J_\L$. Therefore, the leading non-vanishing contribution will come from $\left(\Shell_{OO'O''}^{222}\right)_{(0)_{\rm legs}}$ in special 
\ba
&\left(\Shell^{222}_{\d^2\d^2\d^2}\right)_{(0)_{\rm legs}} (\vk,\vk',\vk'') =  8 \diracpi(\vk+\vk'+\vk'') \int_{\vp} P_{\rm shell}(p) P_{\rm shell}(|\vk-\vp|) P_{\rm shell}(|\vk+\vk'-\vp|) \,,  \label{eq:Sd2d2d2}  
\ea
since terms involving an operator $\d$ will only contribute to higher-derivative operators, because they will contain $F_2(\vp, \vk-\vp) \propto k^2/p^2$ for $p\gg k$. 
Contributions involving $\G_2$ are similarly suppressed. We find then that the only term that contributes to the running of leading-in-derivative zero-leg coefficients is $\Shell^{222}_{\d^2\d^2\d^2}$.

\paragraph{Contributions from $\left(\Shell_{OO'O''}^{222}\right)_{(1)_{\rm legs}}$.}

The one-leg contributions that are relevant for leading-in-derivative terms are
\ba
&\left(\Shell^{222}_{\d^2\d^2\d^2}\right)_{(1)_{\rm legs}} (\vk,\vk',\vk'') =  32 \dlin_\L(\vk+\vk'+\vk'')  \label{eq:Sd2d2d2oneleg}\\ & \hspace{3cm} \int_{\vp} F_2(-\vp, \vk + \vk' + \vk'')P_{\rm shell}(p) P_{\rm shell}(|\vk-\vp|) P_{\rm shell}(|\vk+\vk'-\vp|) \,,  \nonumber \\
&\left(\Shell^{222}_{\d^3\d^2\d^2}\right)_{(1)_{\rm legs}} (\vk,\vk',\vk'') =  24  \dlin_\L(\vk+\vk'+\vk'')  \label{eq:Sd3d2d2oneleg}\\ & \hspace{3cm} \int_{\vp} P_{\rm shell}(p) P_{\rm shell}(|\vk-\vp|) P_{\rm shell}(|\vk+\vk'-\vp|) \,,  \nonumber \\
&\left(\Shell^{222}_{(\G_2\d)\d^2\d^2}\right)_{(1)_{\rm legs}} (\vk,\vk',\vk'') =  16 \dlin_\L(\vk+\vk'+\vk'')  \label{eq:SG2dd2d2oneleg}\\ & \hspace{3cm} \int_{\vp} \s^2_{-\vp,\vk+\vk'+\vk''} P_{\rm shell}(p) P_{\rm shell}(|\vk-\vp|) P_{\rm shell}(|\vk+\vk'-\vp|) \,.  \nonumber
\ea

\section{On the general structure of the shell contributions at the \texorpdfstring{$n$}{n}-point function level} \label{app:doubleshell}

This appendix is dedicated to understand the general structure of the sourcing terms $\Shell_{O_1 O_2\dots O_\o}$ at different orders in $J$. For that, we take the approach of calculating the corrections to $n$-point functions \footnote{An analysis at the level of the partition function leads to intricate momenta structure of the shell terms, such that the $n$-point functions is the easiest way to understand it.}. The main conclusions of this appendix can be summarized as:
The terms of the type $\Shell_{O_1 O_2\dots O_\o}^{\dots 1 \dots }$ lead to nonlocal interactions in the partition function. However, in the limit when external momenta are much smaller than the scale $\L$, these terms vanish due to their kinematic structure. In this limit,  the leading contribution comes from  the terms of the type $\Shell_{O_1 O_2\dots O\o}^{2 2 \dots 2}$.

As a warm up, we start in \refapp{J2structure} by calculating the simplest case of the corrections from $\Shell_{O'O''}$ to $J^2$ operators. Later in \refapp{mostgeneralstruct} we generalize those results to different powers of $J^m$ via $\Shell_{O_1 O_2\dots O_\o}$.

\subsection{The contribution of \texorpdfstring{$\Shell_{OO'}$}{SOO} to \texorpdfstring{$J^2$}{J} terms} \label{app:J2structure} 

The focus of this part is to understand how the $\Shell_{O'O''}$ operators source stochastic contributions (see \reffig{SOOdiag} for the diagrams). As one can see from \refeq{corr2} and \refeq{generalcorr}, $\Shell_{O'O''}$ starts to contribute at order $J^2$, for $P_{\eps,O} = C^{(m)}_O$.
Differently than $\Shell^{2}$, those integrals are not proportional to $\int_{\vp} P_{\rm shell}$, but they involve a more complex momenta structure [e.g., compare \refeq{S2d2calc} to \refeq{S11d2d2}]. 
As we will see, the leading contribution comes from $\Shell^{22}$, since $\Shell^{11}$ and $\Shell^{13}$ are zero due to their kinematic structure. 

To illustrate the suppression of $\Shell^{11}$ and $\Shell^{13}$ and the non-zero contribution from  $\Shell^{22}$, we consider contributions of those terms to $P_{\eps,\d^2}$. Since the $P_{\eps,\d^2}$ term appears at tree-level in the trispectrum level, we will consider the 4-point function for the matching of this term and its correction.
Let us restrict \refeq{ZofJ} to the relevant terms in Fourier space:
\ba
\Z[J_{\L}] = \int \Del\dlin_\L \P[\dlin_\L] \exp&\left(\int_{\vx} J_{\L}(\vx) \ddet(\vx) 
+ P_{\eps,\d^2}(\L) \int_{\vx} J^2_\L(\vx) \left[ \d^2[\dlin_\L](\vx) - \<\d^2\> \right]
  \right)\,, \label{eq:Z_appB} 
\ea
where we have defined
\be
\ddet(\vk) \equiv \sum_O b_O(\L) O[\dlin_\L](\vk)\,,
\ee
for convenience.
We now consider the analysis of the the leading contribution that is proportional  to $P_{\eps,\d^2}(\L)$ in the 4-point function in real space:
\ba
\prod_{m=1}^4\frac{\d }{\d J_{\L}(\vx_m)} Z[J_{\L}]\Big|_{J=0}
&\supset 2 P_{\eps,\d^2}(\L) \dirac(\vx_1-\vx_2) \int \Del\dlin_\L \P[\dlin_\L]
\ddet[\dlin_\L](\vx_3)\ddet[\dlin_\L](\vx_4) \vs
& \hspace{5cm}\times\left[ \d^2[\dlin_\L](\vx_1) - \<\d^2\> \right] + {\rm perm.}\vs
&= 4 P_{\eps,\d^2}(\L) \dirac(\vx_1-\vx_2)
\xi_{\d,\rm det}(\vx_1-\vx_3) \xi_{\d,\rm det}({\vx_2}-\vx_4) 
+ {\rm perm.}\,,
\label{eq:4pt_appB}
\ea
where ``perm.'' denotes permutations of the $\vx_i$ coming from the other possible contractions (hereafter, we drop these other permutations, as they behave analogously to the one written), and
\be
\xi_{\d,\rm det}(\v{r}) \equiv \sum_{O} b_O(\L) \< O[\dlin_\L](\vx) \d[\dlin_\L](\vx+\v{r})\>\,.
\ee

\paragraph{The suppression of \texorpdfstring{$\Shell_{OO'}^{11}$}{SOO11}. } 
Let us now consider the $\Shell^{11}_{\d^2\d^2}$ contribution as example. Using that
$\d_{\L'} = \d_\L + \dlinshell$ and Taylor expanding the exponential, we have
\ba
\Z[J_\L] &\supset 2 \left[b_{\d^2}(\L')\right]^2 \int \Del\dlin_\L \P[\dlin_\L] \int \Del\dlinshell \P[\dlinshell] 
\, \exp \left(\int_{\vx_a} J_{\L}(\vx_a) \ddet(\vx_a)  \right)
\vs
& 
\hspace{1cm}\times \left[ \int_{\vx} J_\L(\vx) \dlin_\L(\vx) \dlinshell(\vx)
\int_{\vx'} J_\L(\vx') \dlin_\L(\vx') \dlinshell(\vx') \right]. \label{eq:Zstep1}
\ea
Integrating out $\dlinshell$, and defining
\be
\xi_{\rm shell}(\vx-\vx') =
\int \Del\dlinshell \P[\dlinshell] \dlinshell(\vx) \dlinshell(\vx'),
\ee
we obtain
\ba
\Z[J_\L] & \supset 2 \left[b_{\d^2}(\L')\right]^2 \int \Del\dlin_\L \P[\dlin_\L] \, \exp \left(\int_{\vx_a} J_{\L}(\vx_a) \ddet(\vx_a)  \right) \vs
& \quad
\times \int_{\vx} J_\L(\vx) \dlin_\L(\vx) 
\int_{\vx'} J_\L(\vx') \dlin_\L(\vx') \xi_{\rm shell}(\vx-\vx') .
\label{eq:S11_appB}
\ea
Comparing with \refeq{Z_appB}, evaluated at leading order $\d[\dlin_\L] \to \dlin_\L$, we see that \refeq{S11_appB} has a similar structure to the $P_{\eps,\d^2}$ contribution, but with $\dirac(\vx-\vx')$ replaced with $\xi_{\rm shell}(\vx-\vx')$. That is, \refeq{S11_appB} is \emph{nonlocal}, while the interaction written in the partition function is local. 
We obtain the contribution to the 4-point function which corresponds to \refeq{4pt_appB}:
\footnote{Notice we can use the absence of preferred directions in the initial conditions, and the assumption of a spherical filter $W_\L$, to write
\ba
\xi_{\rm shell}(\vx-\vx') = \xi_{\rm shell}(|\vx-\vx'|) &=
\int_{\vk\in \rm shell} \Plin(k) \exp[\ii \vk\cdot (\vx-\vx')] \vs
&= \frac1{2\pi^2} \L^2 \lambda \Plin(\L) \frac{\sin (\L r)}{\L r}\Big|_{r = |\vx-\vx'|}
= \sigma_{\rm shell}^2 \frac{\sin (\L r)}{\L r}\Big|_{r = |\vx-\vx'|}\, .
\ea
In the second line, we have additionally assumed a shell of infinitesimal width $\lambda$. This shows the expected structure: $\xi_{\rm shell}(0)=\sigma_{\rm shell}^2$, and it decays on the scale $r \sim 1/\L$.}
\ba
\prod_{m=1}^4\frac{\d }{\d J_{\L}(\vx_m)} Z[J_{\L}]\Big|_{J=0}
&\supset 4 \left[b_{\d^2}(\L')\right]^2 \xi_{\rm shell}(|\vx_1-\vx_2|)
\xi_{\d,\rm det}(\vx_1-\vx_3) \xi_{\d,\rm det}({\vx_2}-\vx_4) \label{eq:4pt_appB2} \\
&\quad + 4 \left[b_{\d^2}(\L')\right]^2 \xi_{\rm shell}(|\vx_1-\vx_2|)
\xi_{\d,\rm det}(\vx_2-\vx_3) \xi_{\d,\rm det}({\vx_1}-\vx_4) 
 \,. \nonumber
\ea

In order to connect \refeq{4pt_appB} and \refeq{4pt_appB2}, we now assume that
the 4-point function is integrated against a test function $f(\vx_1,\vx_2,\vx_3,\vx_4)$ which defines the configurations or quadrilateral bins for which the 4-point function is measured. If we asssume that the current $J_\L(\vk)$ only has support on large scales, $k \lesssim k_{\rm max} \ll \Lambda$,  \refeq{4pt_appB} straightforwardly yields
\ba
&\int_{\vx_1,\ldots \vx_4} f(\vx_1,\ldots \vx_4)
\prod_{m=1}^4\frac{\d }{\d J_{\L}(\vx_m)} Z[J_{\L}]\Big|_{J=0} \vs
&\hspace{2cm}= 4 P_{\eps,\d^2}(\L) \int_{\vx_1,\vx_3,\vx_4}\!\!\!\! f(\vx_1,\vx_1,\vx_3,\vx_4)
\xi_{\d,\rm det}(\vx_1-\vx_3) \xi_{\d,\rm det}({\vx_2}-\vx_4)  \,.
\ea
On the other hand, \refeq{4pt_appB2} yields, after Fourier transforming,
\ba
& 4 \left[b_{\d^2}(\L')\right]^2 \int_{\vx_1,\vx_2,\vx_3,\vx_4}\!\!\!\!
   \xi_{\d,\rm det}(\vx_1-\vx_3)\xi_{\d,\rm det}(\vx_2-\vx_4) \xi_{\rm shell}(\vx_1-\vx_2) f(\vx_1,\vx_2,\vx_3,\vx_4) 
 \vs
  & = 4 \left[b_{\d^2}(\L')\right]^2 \int_{\vk_1,\vk_2,\vk_3,\vk_4}\!\!\!\!
   P_{\d,\rm det}(\vk_3) P_{\d,\rm det}(\vk_4) P_{\rm shell}(\vk_2+\vk_4) \tilde{f}(\vk_1,\vk_2,\vk_3,\vk_4)\diracpi(\vk_{1234}) \label{eq:S11realspace}  \,, 
\ea
in which $\tilde{f}$ is the Fourier transform of $f$. Notice that this term is zero after using that $|\vk_2+\vk_4| < \L$, after taking $|\vk_2| \ll \L$ and $|\vk_4| \ll \L$ and using the support region of $f$. It is straightforward to generalize this argument to other  $\Shell^{11}_{OO'}$.

\paragraph{The suppression of \texorpdfstring{$\Shell_{OO'}^{13}$}{SOO13}.}
The $\Shell_{OO'}^{13}$ corrections correspond to an $\Shell_{OO'}^{11}$ term with an extra loop in one of the vertices (see \reffig{SOOdiag}). 
Similarly to \refeq{Zstep1}, we can write, taking as example the $O = \d^2, O' = \d^4$ contribution 
\ba
\Z[J_\L] &\supset  b_{\d^2}(\L')b_{\d^4}(\L') \int \Del\dlin_\L \P[\dlin_\L] \int \Del\dlinshell \P[\dlinshell] 
{\, \exp \left(\int_{\vx_a} J_{\L}(\vx_a) \ddet(\vx_a)  \right)}
\vs
& 
\hspace{2.5cm}\times \left[ \int_{\vx} J_\L(\vx) \dlin_\L(\vx) \dlinshell(\vx)
\int_{\vx'} J_\L(\vx') \dlin_\L(\vx') \left( \dlinshell(\vx') \right)^3 \right]\,,
\ea
such that after integrating out $\dlinshell$ we obtain
\ba
\Z[J_\L] & \supset  \left[b_{\d^2}(\L')\right]^2 \xi_{\rm shell}(0) \int \Del\dlin_\L \P[\dlin_\L] {\, \exp \left(\int_{\vx_a} J_{\L}(\vx_a) \ddet(\vx_a)  \right)} \vs
& \hspace{4cm}
 \int_{\vx} J_\L(\vx) \dlin_\L(\vx) 
\int_{\vx'} J_\L(\vx') \dlin_\L(\vx') \xi_{\rm shell}(\vx-\vx')\,,
\ea
to write
\ba
 & 4 \left[b_{\d^2}(\L')\right]^2 \xi_{\rm shell}(0) \int_{\vk_1,\vk_2,\vk_3,\vk_4} 
P_{\d,\rm det}(\vk_3) P_{\d,\rm det}(\vk_4)
  P_{\rm shell}(\vk_2+\vk_4) \tilde{f}(\vk_1,\vk_2,\vk_3,\vk_4)\diracpi(\vk_{1234}) \label{eq:S13realspace}  \,,
\ea
which, similarly to \refeq{S11realspace}, is zero. Since all contributions to the operators appearing in the bias expansion can be constructed from local products in real space of gradients of the gravitational potential, and ultimately $(\partial_i/\nabla^2)\dlin$, \refeq{S13realspace} is generic for all operators $\Shell_{OO'}^{13}$.

\paragraph{The \texorpdfstring{$\Shell_{OO'}^{22}$}{SOO22} contributions.}
Differently than  $\left(\Shell^{11}_{\d^2\d^2}\right)_{(0)_{\rm leg}}$ and $\left(\Shell^{13}_{\d^2\d^2}\right)_{(1)_{\rm leg}}$ which are zero since they are proportional to an external momenta $P_{\rm shell}(k)$, $\Shell_{OO'}^{22}$ is not zero due to the integrated shell structure [see e.g. \refeq{S22_d2d2_calc}]. 
This term will therefore lead to corrections to $P_{\eps,\unit}$ and $P_{\eps,\d}$ (see \refsec{SOOmain}).
For comparison with the previous two sections, we consider corrections to $P_{\eps,\d^2}$.
Following \refeq{Zstep1}, the $\Shell_{OO'}^{22}$ (see \reffig{SOOdiag}) corrections can be written, taking as example the $O = \d^3, O' = \d^3$ contribution, as
\ba
\Z[J_\L] &\supset  \left[b_{\d^3}(\L')\right]^2 \int \Del\dlin_\L \P[\dlin_\L] \int \Del\dlinshell \P[\dlinshell] 
{\, \exp \left(\int_{\vx_a} J_{\L}(\vx_a) \ddet(\vx_a)  \right)}
\vs
& 
\hspace{2cm}\times \left[ \int_{\vx} J_\L(\vx) \dlin_\L(\vx) \left( \dlinshell(\vx) \right)^2
\int_{\vx'} J_\L(\vx') \dlin_\L(\vx') \left( \dlinshell(\vx') \right)^2 \right]\,,
\ea
such that after integrating out $\dlinshell$ we obtain
\ba
\Z[J_\L] & \supset  \left[b_{\d^2}(\L')\right]^2 \int \Del\dlin_\L \P[\dlin_\L] {\, \exp \left(\int_{\vx_a} J_{\L}(\vx_a) \ddet(\vx_a)  \right)} \vs
& \hspace{2cm}
 \int_{\vx} J_\L(\vx) \dlin_\L(\vx) 
\int_{\vx'} J_\L(\vx') \dlin_\L(\vx') \left[\xi_{\rm shell}(\vx-\vx')\right]^2\,,
\ea
to write
\ba
 &  \left[b_{\d^2}(\L')\right]^2  \int_{\vk_1,\vk_2,\vk_3,\vk_4} \left [\diracpi(\vk_{1234}) P_{\d,\rm det}(\vk_3) P_{\d,\rm det}(\vk_4)\tilde{f}(\vk_1,\vk_2,\vk_3,\vk_4)  \right. \vs 
 & \hspace{5cm}\left. \int_{\vk_{\rm shell}} P_{\rm shell}(\vk_{\rm shell}) P_{\rm shell}(\vk_2+\vk_4-\vk_{\rm shell}) \right]   \,, \label{eq:S22realspace} 
\ea
which, differently than \refeq{S11realspace} and \refeq{S13realspace} is not zero since $\left[\xi_{\rm shell}(x)\right]^2$  will lead to a convolution in Fourier space. Again, this can be generalized to other operators by noting that they can be written as real-space products of gradients of the gravitational potential.

\subsection{The structure of the \texorpdfstring{$n$}{n}-point functions for the most general case} \label{app:mostgeneralstruct}

The previous section focused on the contributions of $(J^1)^2$ terms to $J^2$ terms via $\Shell_{OO'}$. When calculating the terms for \refeq{generalcorr}, the conclusions of the former section have to be generalized to include: first, operators from different powers in $J$ contributing together (e.g. operators $J^1$ and $J^2$ contributing to $J^3$) and second, by considering the most generic case $\Shell_{O_1 \dots O_\o}$ and not simply $\Shell_{OO'}$.   

\paragraph{Contributions with different powers in $J$.} 
We start by considering the example of a term $J^1$ and another $J^2$ contributing to a $J^3$ term via $\Shell_{OO'}$. 
For the $J^3$ term $B_{\eps,\d^2}(\L)$, the 5-point function in real space 
\ba
&\prod_{m=1}^5\frac{\d }{\d J_{\L}(\vx_m)} Z[J_{\L}]\Big|_{J=0}
\supset B_{\eps,\d^2}(\L) \dirac(\vx_1-\vx_3)\dirac(\vx_1-\vx_2)  
\vs
& \hspace{2cm}\times \int \Del\dlin_\L \P[\dlin_\L]
\ddet[\dlin_\L](\vx_4)\ddet[\dlin_\L](\vx_5)\left[ \d^2[\dlin_\L](\vx_1) - \<\d^2\> \right] 
\vs
&\hspace{1cm} = B_{\eps,\d^2}(\L) \dirac(\vx_1-\vx_2) \dirac(\vx_1-\vx_3)
\xi_{\d,\rm det}(\vx_1-\vx_4) \xi_{\d,\rm det}(\vx_3-\vx_5) \,,
\ea
contains $B_{\eps,\d^2}(\L)$ at tree level. We obtain that the $\Shell_{OO'}^{11}$ correction to this term is
\ba
\prod_{m=1}^5\frac{\d }{\d J_{\L}(\vx_m)} Z[J_{\L}]\Big|_{J=0} & \supset  \left[b_{\d^2}(\L')\right]\left[P_{\eps,\d^2}(\L')\right]\dirac(\vx_1-\vx_2)  \\ 
& \quad \xi_{\rm shell}(\vx_1-\vx_3)
\xi_{\d,\rm det}(\vx_1-\vx_4) \xi_{\d,\rm det}(\vx_3-\vx_5)  \,. \nonumber
\ea
Integrating the 5-point function against a test function $f(\vx_1,\vx_2,\vx_3,\vx_4,\vx_5)$ leads to
\ba
&\int_{\vx_1,\ldots \vx_5} f(\vx_1,\ldots, \vx_5)
\prod_{m=1}^5\frac{\d }{\d J_{\L}(\vx_m)} Z[J_{\L}]\Big|_{J=0} \vs
&\quad = B_{\eps,\d^2}(\L) \int_{\vx_1,\vx_4,\vx_5}\!\!\!\! f(\vx_1,\vx_1,\vx_1,\vx_4,\vx_5)
\xi_{\d,\rm det}(\vx_1-\vx_4) \xi_{\d,\rm det}(\vx_3-\vx_5)  \,.
\ea
On the other hand, the $\Shell_{OO'}^{11}$ correction yields
\ba
& \left[b_{\d^2}(\L')\right]\left[P_{\eps,\d^2}(\L')\right] \int_{\vx_1,\vx_3,\vx_4,\vx_5}\!\!\!\!
   \xi_{\d,\rm det}(\vx_1-\vx_4)
 \vs
 & \hspace{4cm}\, \xi_{\d,\rm det}(\vx_3-\vx_5) \xi_{\rm shell}(\vx_1-\vx_3) f(\vx_1,\vx_1,\vx_3,\vx_4,\vx_5)   \,. 
  \vs
  & \quad = \left[b_{\d^2}(\L')\right]\left[P_{\eps,\d^2}(\L')\right] \int_{\vk_1,\vk_2,\vk_3,\vk_4,\vk_5}\!\!\!\!
   P_{\d,\rm det}(\vk_4)P_{\d,\rm det}(\vk_5)
 \vs
 & \hspace{4cm}\, P_{\rm shell}(\vk_3+\vk_5) \tilde{f}(\vk_1,\vk_2,\vk_3,\vk_4,\vk_5)\diracpi(\vk_{12345})  \,, 
\ea
which is zero using that $|\vk_3+\vk_5| < \L$, after taking $|\vk_3| \ll \L$ and $|\vk_5| \ll \L$. 
The $\Shell_{OO'}^{22}$ correction yields
\ba
 &  \left[b_{\d^2}(\L')\right]\left[P_{\eps,\d^2}(\L')\right]  \int_{\vk_1,\vk_2,\vk_3,\vk_4,\vk_5} \left [\diracpi(\vk_{12345}) P_{\d,\rm det}(\vk_4) P_{\d,\rm det}(\vk_5)\tilde{f}(\vk_1,\vk_2,\vk_3,\vk_4,\vk_5)  \right. \vs 
 & \hspace{5cm}\left. \int_{\vk_{\rm shell}} P_{\rm shell}(\vk_{\rm shell}) P_{\rm shell}(\vk_2+\vk_4-\vk_{\rm shell}) \right]   \,, \label{eq:S22realspace_cross} 
\ea
which is not zero due to the convolution in Fourier space.

\paragraph{The general structure of \texorpdfstring{$\Shell_{O_1 \dots O_\o}$}{SO...O}.} 
One can use a very similar argument to the one presented in \refapp{J2structure} to show that {\it any shell contribution of the type $\Shell_{O_1 O_2\dots O_\o}^{\dots 1 \dots }$ is zero in the limit of soft external momenta}. Considering whatever $n$-point function contains this term at tree level, similarly to \refeq{S11realspace} and \refeq{S13realspace}, this term will contain the structure 
\ba
& \int d^3\v{r}\, \xi_{\rm shell}(r) f(\vx_1,\vx_1+\v{r},\dots)  \,, \label{eq:S11111general}
\ea
in the operator argument that has one single shell expanded. Again, due to the orthogonality of $\xi_{\rm shell}$ and $f$, this term is zero.
In other words, every contribution that involves one of the operators expanded with a single shell will be zero, which as a consequence (and noticing that other contributions are higher-loop) leads to the conclusion that {\it the only non-zero contribution comes from   $\Shell_{O_1 O_2\dots O\o}^{2 2 \dots 2}$}.

\bibliographystyle{JHEP}
\bibliography{main}

\end{document}